\documentclass[12pt,preprint]{aastex}

\shorttitle{Unveiling the core of the Globular Cluster M\,15 in the
  Ultraviolet} 
\shortauthors{Dieball et al.}

\begin{document}

\title{Unveiling the core of the Globular Cluster M\,15 in the
  Ultraviolet\footnote{Based on observations with the NASA/ESA
    {\it Hubble Space Telescope}, obtained at the Space Telescope Science
    Institute, which is operated by the Association of Universities
    for Research in Astronomy, Inc., under NASA contract NAS5-26555.}} 

\author{A. Dieball, C. Knigge}
\affil{Department of Physics and Astronomy, University Southampton,
  SO17 1BJ, UK}

\author{D. R. Zurek, M. M. Shara}
\affil{Department of Astrophysics, American Museum of Natural History,
New York, NY 10024}

\author{K. S. Long}
\affil{Space Telescope Science Institute, Baltimore, MD 21218}

\author{P. A. Charles}
\affil{South African Astronomical Observatory, PO Box 9, Observatory,
  7935, South Africa}

\and

\author{D. Hannikainen}
\affil{University of Helsinki, P.O. Box 14, SF-00014 Helsinki, Finland}

\begin{abstract}

We have obtained deep FUV and NUV images of the inner region of the
dense globular cluster M\,15 with the HST ACS.  
The FUV--NUV colour-magnitude diagram shows a well-defined
track of horizontal branch stars, as well as a trail of blue
stragglers and white dwarfs. The main-sequence turnoff is clearly
visible at FUV $\simeq 23.5$ mag and FUV--NUV$ \simeq 3$ mag, and the
main-sequence stars form a prominent track that extends at least
2 mag below the main-sequence turnoff. As such, this is the
deepest FUV--NUV colour-magnitude diagram of a globular cluster
presented so far. Cataclysmic variable and blue straggler candidates
are the most centrally concentrated stellar populations, which might
either be an effect of mass segregation or reflect the preferred
birthplace in the dense cluster core of such dynamically-formed
objects. 
We find 41 FUV sources that exhibit significant variability. We
classify the variables based on an analysis of their UV colours and
variability properties. We find four previously known RR Lyrae and 13
further RR Lyrae candidates, one known Cepheid and six further
candidates, six cataclysmic variable  candidates, one known and one
probable SX Phoenicis star, and the well known low-mass X-ray binary
AC\,211. Our analysis represents the first detection of SX Phoenicis
pulsations in the FUV. We find that Cepheids, RR Lyrae stars and SX
Phoenicis exhibit massive variability amplitudes in this wave band
(several magnitudes).    

\end{abstract}


\section{Introduction}

Globular clusters (GCs) in the Milky Way are old, gravitationally bound stellar
systems. Stellar densities in their cores can be extremely high,
reaching up to $10^{6}\ {\rm stars}/{\rm pc}^{3}$, hence making them
ideal crash-test laboratories for studying the dynamics in such dense
environments. Close encounters and even direct collisions with
resulting mergers between the cluster stars are inevitable, leading to
a variety of dynamically formed, exotic stellar populations like blue
stragglers (BSs), cataclysmic variables (CVs), low-mass X-ray binaries
(LMXBs), and probably binary horizontal branch (HB) stars and other
close binary (CB) systems. CBs are important for our understanding of
GC evolution, since the binding energy of a few CBs can rival that of
a GC. Thus, by transferring their orbital energy to
passing single stars, CBs can significantly affect the dynamical
evolution of the cluster (e.g.\ Elson et al.\ 1987, Hut et al.\ 1992,
and references therein). This depends critically on the number of CBs;
if there are only a few CBs, long-term interactions dominate the cluster
evolution, but the presence of many CBs can lead to violent
interactions that heat the cluster and cause its expansion and
evaporation on significantly shorter time-scales. 
Therefore, knowledge of the CB populations in the cores of clusters
is crucial to understanding the evolution of clusters.

However, despite their impact on cluster evolution and their
importance for our understanding of dynamical binary formation and
binary evolution, there have been only a few detections of binaries in
GCs during the past decades. The study of binaries has been
extremely difficult due to the spatial resolutions and detection
limits of the available telescopes. Only with the arrival of {\it Chandra}
and the {\it Hubble Space Telescope (HST)}, 
with their improved imaging capabilities, has it been possible to finally
detect large numbers of CVs and other binary systems (e.g.\ Grindlay
et al.\ 2001; Edmonds et al.\ 2003a, 2003b; Knigge et al.\ 2002, 2003;
Dieball et al.\ 2005a and references therein).  

CBs and other dynamically formed systems share an observational
characteristic: they show a spectral energy distribution that is bluer
than that of other cluster members and emit radiation in the 
far-ultraviolet (FUV). ``Ordinary'' stars, like main-sequence (MS)
stars and red giants (RGs), are too cool to show up at these
wavelengths. As a consequence, crowding is generally not a severe
problem in FUV imaging, very much unlike optical imaging, which is
often nearly impossible in dense cluster cores. Thus, a deep,
high-resolution FUV-survey is the ideal tool to detect and study
these stellar species.   

So far, deep FUV studies have been carried out in detail only for two
clusters: 47\,Tuc \citep{knigge1,knigge2} and NGC\,2808 (Brown et
al. 2001; Dieball et al. 2005). In 47\,Tuc, Knigge et al.\ (2002)
found FUV counterparts to all of the {\it Chandra} CV candidates known at
that time in the core of this cluster (Grindlay et
al. 2001)\footnote{Note that recently \citet{heinke2005} published a
  catalog of X-ray sources of 47\,Tuc comprising 300 objects within
  one half-mass radius. Work is in progress to identify counterparts
  to all the sources located within the FUV field of view (C. Knigge
  2006, private communication).} and
confirmed their CV nature. In addition, several additional CV
candidates were suggested, and a clean, well-populated FUV BS
sequence and numerous young white dwarfs (WDs) on the upper end of the
cooling sequence were found. In NGC\,2808, Brown et al.\ (2001)
used FUV and NUV imaging to uncover a population of sub-luminous
hot HB stars that probably underwent a late helium-core flash on the
WD cooling curve. Our own reanalysis of this data set pushed
the detection limits deeper and revealed numerous BSs, CV candidates,
and hot young WDs in this cluster \citep{dieball1}.

M\,15 is one of the oldest (13.2 Gyr; McNamara et al.~2004)
low-metallicity ($\rm{[Fe/H]} = -2.26$; Harris 1996) GCs in our
Galaxy, at a distance of 10.3 kpc (van den Bosch et al.~2006). It has
received considerable attention in the literature for numerous
reasons. It is probably a core-collapsed cluster, with a very small
and very dense core ($\rm{R}_{c} \approx 0\farcm07$, 
$\rho_{c} \approx 7 \times 10^{6} M_\odot {\rm pc}^{-3}$;
van den Bosch et al.~2006). 
Because of its unusual compact core, high central velocity
dispersion, and luminosity profile, it has even been suggested that
M\,15 hosts an intermediate-mass black hole (IMBH) of up to a few
thousand solar masses (e.g.\ Peterson et al.~1989; Guhathakurta et
al.~1996; Gebhardt et al.~1997; Gerssen et al.~2002). However,
Baumgardt et al.\ (2003) presented $N$-body models that are capable of
explaining the observed velocity dispersion and luminosity profile
without the need for a central IMBH, but with a concentration of
neutron stars (NSs) and massive WDs in the centre due
to mass segregation. McNamara et al.\ (2003) also found that there is
little evidence for an IMBH, their proper-motion dispersion profile of
the inner region of M\,15 being in good agreement with the predictions
from Baumgardt et al.\ (2003). Most recently, van\,den\,Bosch et
al.\ (2006) suggested a mass of 3400 $M_\odot$ in the central 0.05 pc,
which could either be an IMBH, or a large number of compact objects,
or a combination of both.  
 
M\,15 is also the only Galactic GC known to harbour {\it two} bright
LMXBs. The first one, 4U\,2127+119, was identified by Auri{\`e}re et
al. (1984) with the optical counterpart AC\,211. It has an orbital
period of 17.1 hr (Ilovaisky et al.~1993) and is one of the optically
brightest LMXBs known. Its high $L_{\rm{opt}}/L_{\rm{X}}$ ratio
suggests that the system is an accretion disk corona source. White \&
Angelini (2001) used {\it Chandra} observations to
resolve 4U\,2127+119 into {\it two} X-ray sources, separated by just
$2\farcs7$ (see also Charles et al.\ 2002; Hannikainen et al.\
2005). One of these was the previously known LMXB AC\,211. The 
second source, CXO\,J212958.1+121002 or M\,15 X-2, is actually 2.5
times brighter than AC\,211 in X-rays and is the likely source of the
X-ray bursts observed in 1988 and 2000 (Dotani et al.~1990; Smale
2001). Its optical counterpart is a blue $U \approx 19$ mag
star (star 590 in De\,Marchi \& Paresce 1994). Dieball et al.\ (2005b)
found that M\,15 X-2 is an ultra-compact X-ray binary (UCXB) with an
orbital period of 22.6 minutes, only the third confirmed such object in a
GC. 
 
M\,15 is one of the three original Oosterhoff type II clusters
identified in \citet{oosterhoff}. More than 200 variable stars are
known in this cluster; the majority of them are RR Lyrae stars (e.g.\
Silbermann \& Smith 1995; Clement et al.\ 2001; Arellano et al.\
2006). \citet{clement01} list 204 entries in their catalog
of variable stars in M\,15, out of which 149 are RR Lyrae stars (at
least 55 are of type RRab, 52 of type RRc, and 23 of type RRd).
\citet{tuairisg} classified 19 RRc (10 new ones), 12 RRab (six
previously unknown) and two RRd (one previously unknown). However, the
classification of the RR Lyrae subtype is not always consistent with 
\citet{clement01}.
 
The cluster also hosts other interesting stellar populations, among
them a large number of BSs (e.g.\ Yanni et al.~1994). At least
three SX Phoenicis stars (Jeon et al.\ 2001; Kravtsov \& Zheleznyak
2003), eight millisecond pulsars (Phinney 1996) and three dwarf nova (DN)
candidates are known (Shara et al.~2004; Hannikainen et al.~2005). De
Marchi \& Paresce (1994, 1996) also found extreme blue stars in their
broadband $UV$ colour-magnitude diagram (CMD), which were  explained by
D'Antona et al.\ (1995) as the merger product of two low-mass helium
white dwarfs. Most of the above mentioned exotica are indicators of
strong dynamical interaction. M\,15's optical CMD (Van der Marel et
al.~2002) shows a prominent blue horizontal branch (BHB) and extreme
blue horizontal branch (EHB) stars with a few gaps along the HB (see
below our Fig.~\ref{cmd_opt}, right panel).

Here we present FUV and NUV data of M\,15 taken with the Advanced
Camera for Surveys (ACS) on board the {\it HST}. In Sect.~\ref{data}
we describe the data and their reduction. We present the analysis of
the photometry and the FUV versus NUV CMD in Sect.~\ref{cmd}. Our
search for variability among our catalog stars and a study of the
variable sources are given in Sect.~\ref{variability}. Our results
are summarized in Sect.~\ref{summary}.   

\section{Observations and data reduction}
\label{data}

M15 was observed with the ACS onboard {\it HST} for a total of 10
orbits in 2003 October and 2004 October to December. The observations
were spaced out through this period in order to improve our chances of
measuring time-variable phenomena. On the first visit in 2003 October,
eight images with individual exposures of 290 s were obtained using
the NUV F220W filter in the High Resolution Channel (HRC). The
remaining visits were used to obtain a total of 90 images using
the FUV F140LP filter in the Solar Blind Channel (SBC).  Each of the
SBC visit consisted of eight 300 s exposures, followed by either a
40 s or a 140 s exposure at the end of the visit. The same pointing
and no dither pattern was used for all of the observations. An
observing log can be found in Table~\ref{obslog}. The pixel sizes are
$0\farcs034 \times 0\farcs030$ for the SBC, resulting in a nominal
field of view of $35\arcsec \times 31\arcsec$, and $0\farcs028 \times
0\farcs025$ pixels for the HRC, covering a nominal $29\arcsec \times
26\arcsec$ field of view. 
 
\begin{table}
\begin{center}
\caption{\label{obslog} Observation log. Col. (1): Visit. Col. (2):
  Data set. Col. (3): Start date of the observations. Col. (4): Total exposure
  time for that data set. Cols. (5) and (6): Detector and Filter,
  respectively. See text for details.} 
\begin{tabular}{llcrcl}
\tableline
visit & dataset    & start date       & total exp.time & detector (ACS) & filter\\
\tableline\tableline
1 & J8SI08010  & 2003-10-27 02:03:00  & 2320 & HRC & F220W\\\tableline
2 & J8SI01011  & 2004-10-14 05:44:00  & 2400 & SBC & F140LP\\ 
2 & J8SI01RKQ  & 2004-10-14 06:29:00  & 40   & SBC & F140LP\\\tableline 
3 & J8SI01021  & 2004-10-14 07:15:00  & 2400 & SBC & F140LP\\ 
3 & J8SI01S8Q  & 2004-10-14 08:00:00  & 140  & SBC & F140LP\\\tableline 
4 & J8SI01031  & 2004-10-14 08:51:00  & 2400 & SBC & F140LP\\ 
4 & J8SI01SIQ  & 2004-10-14 09:36:00  & 140  & SBC & F140LP\\\tableline 
5 & J8SI01041  & 2004-10-14 10:27:00  & 2400 & SBC & F140LP\\ 
5 & J8SI01T4Q  & 2004-10-14 11:12:00  & 140  & SBC & F140LP\\\tableline 
6 & J8SI01051  & 2004-10-14 12:03:00  & 2400 & SBC & F140LP\\ 
6 & J8SI01TEQ  & 2004-10-14 12:48:00  & 140  & SBC & F140LP\\\tableline 
7 & J8SI02011  & 2004-10-24 10:27:00  & 2400 & SBC & F140LP\\ 
7 & J8SI02LTQ  & 2004-10-24 11:13:00  & 40   & SBC & F140LP\\\tableline 
8 & J8SI03011  & 2004-11-06 19:55:00  & 2400 & SBC & F140LP\\ 
8 & J8SI03GCQ  & 2004-11-06 20:40:00  & 40   & SBC & F140LP\\\tableline 
9 & J8SI04011  & 2004-11-17 11:49:00  & 2400 & SBC & F140LP\\ 
9 & J8SI04JDQ  & 2004-11-17 12:34:00  & 40   & SBC & F140LP\\\tableline 
10 & J8SI05011  & 2004-11-26 16:30:00  & 2400 & SBC & F140LP\\ 
10 & J8SI05YYQ  & 2004-11-26 17:16:00  & 40   & SBC & F140LP\\\tableline 
11 & J8SI06011  & 2004-12-05 03:36:00  & 2400 & SBC & F140LP\\ 
11 & J8SI06BSQ  & 2004-12-05 04:22:00  & 40   & SBC & F140LP\\ 
\tableline
\end{tabular}
\end{center}
\end{table}

As the ACS focal surfaces are inclined towards the principal rays, the
resulting exposures are geometrically distorted. This distortion consists
of two effects, namely, a tilted elongation of the ACS apertures,
causing the pixel scale to be smaller along the radial direction than
along the tangential direction of the Optical Telescope Assembly field
of view, and a variation of pixel area across the detector. In order
to correct these effects, each pixel of the distorted original image
has to be mapped onto pixels in a rectified, undistorted output image,
taking into account shifts and rotations between images and the -
known - optical distortion of the camera. For a detailed discussion on
the distortion effects of ACS imaging and the use of {\tt multidrizzle} we
refer the reader to the {\it HST} ACS data handbook and references therein. 

\begin{figure}
\includegraphics[width=16cm]{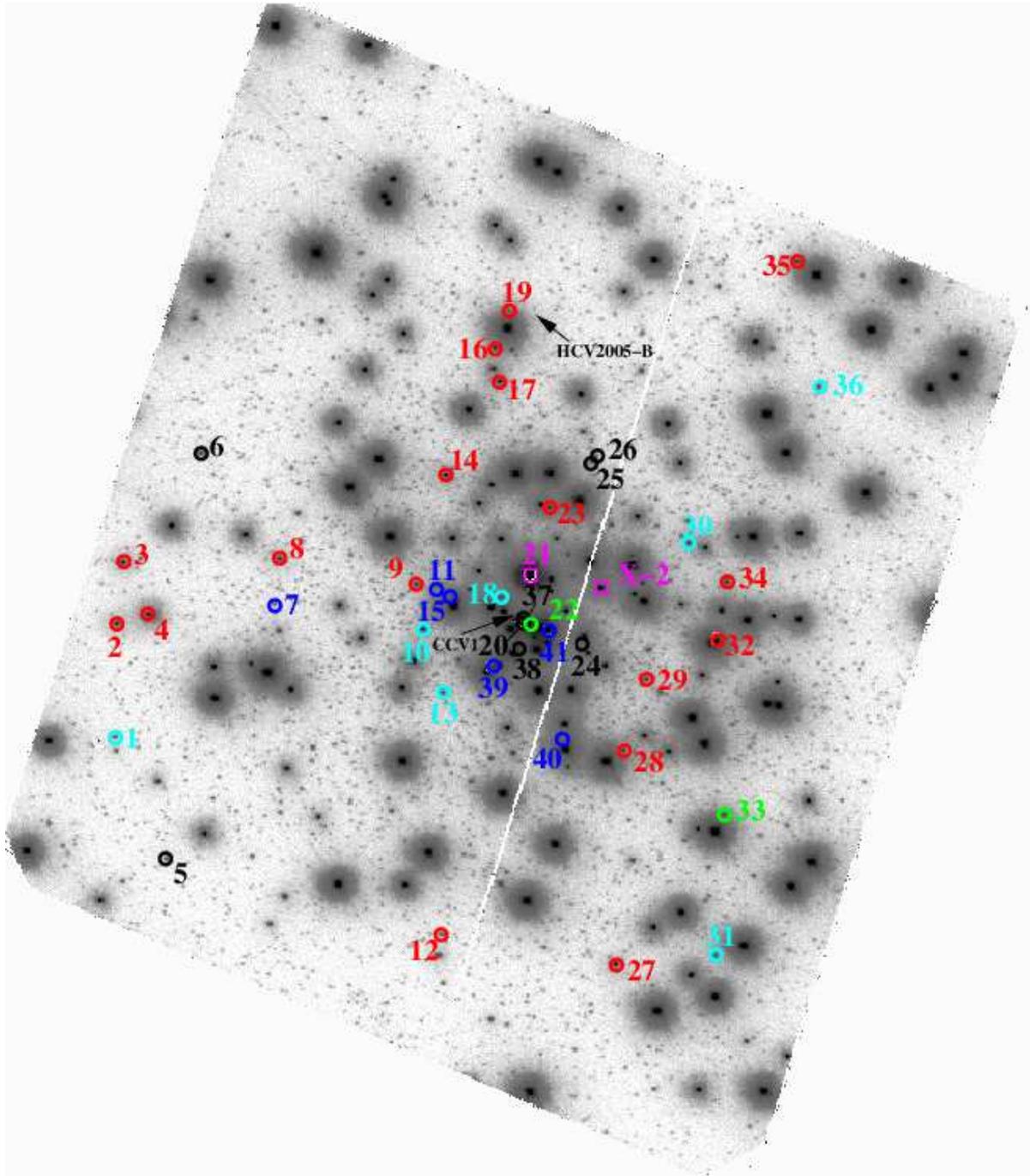}
\caption{\label{image1} Combined and geometrically corrected
  master image of all FUV SBC/F140LP exposures taken from the
  central region of M\,15. North is up and east to the left. The pixel
  scale is $0\farcs025$ pixel$^{-1}$, and the field of view is $35\arcsec
  \times 31\arcsec$. Variables are marked with  
  circles and their number given in
  Tables~\ref{periods1} and \ref{periods2}. The colours denote: red -
  RR Lyrae and candidates, blue - CV candidates, green - SX Phoenicis
  and candidates, cyan - Cepheid and candidates, magenta squares -
  LMXBs, black - other variables. The locations of two further X-ray
  sources, HCV\,2005-A and HCV\,2005-B, which are situated in our field of
  view and are suspected DNe, are indicated with arrows. Note 
  that their FUV counterparts are too faint to show up in our FUV
  master image, but the sources could be identified in the NUV. See
  Sect.~\ref{dn} for details.}       
\end{figure}

\begin{figure}
\includegraphics[width=16cm]{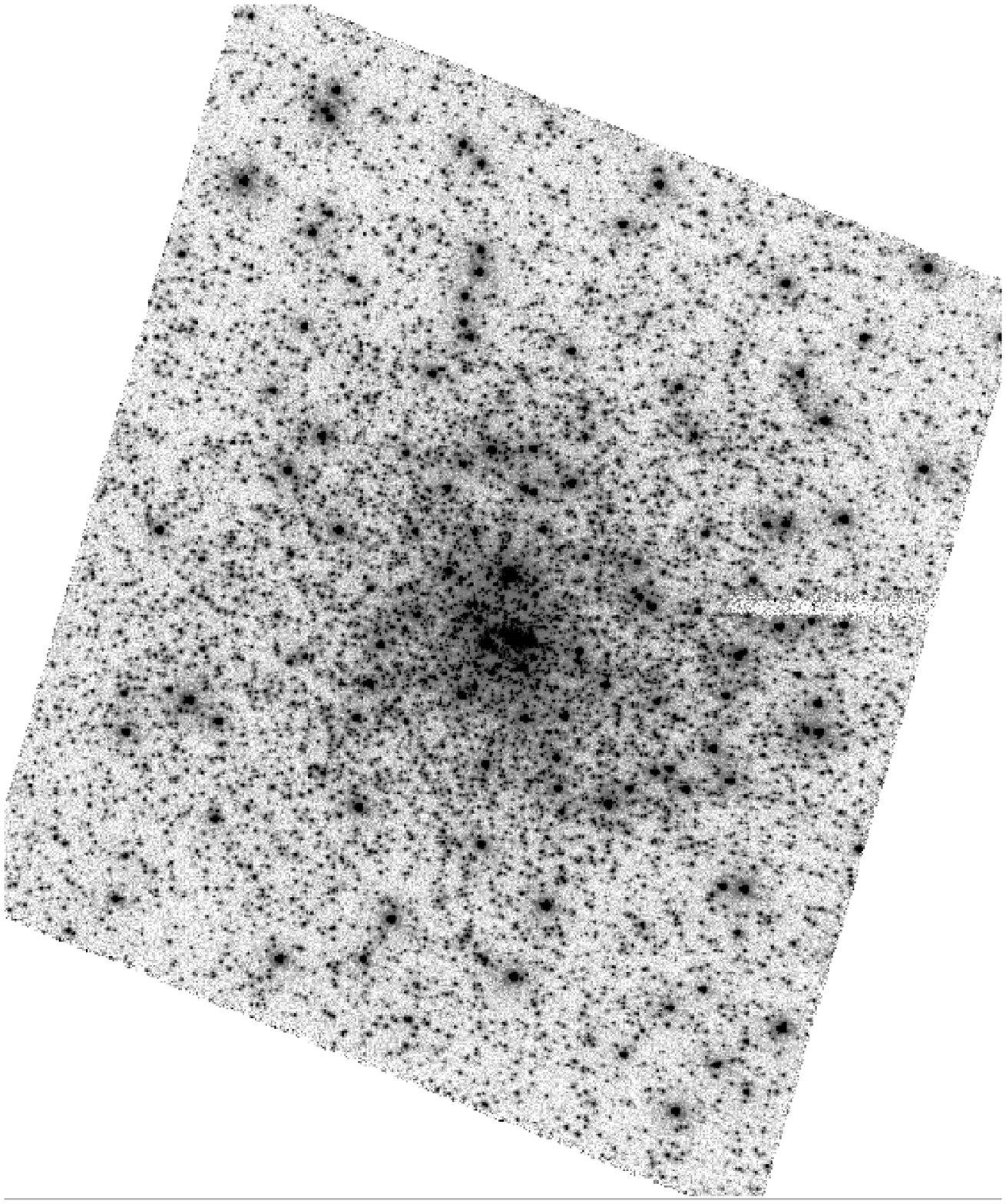}
\caption{\label{image2} Master image of all NUV exposures
  of the same region of M\,15. The orientation is the same as in
  Fig.~\ref{image1}. 
  Both Fig.~\ref{image1} and \ref{image2} are
  displayed on a logarithmic intensity scale in order to bring out the
  fainter sources.}    
\end{figure}

We created master images for each filter using {\tt multidrizzle}
running under {\tt PyRAF}. As a first step, we geometrically corrected
the single flat-fielded images. To account for small shifts between the
desired and actual pointing positions, we used the 
{\tt IRAF}\footnote{{\tt IRAF} (Image  
  Reduction and Analysis Facility) is distributed by the National
  Astronomy and Optical Observatory, which is operated by AURA,
  Inc., under cooperative agreement with the National Science
  Foundation.} 
tasks {\tt crossdriz} and {\tt shiftfind} on the single drizzled
(i.e. geometrically corrected) images. The shifts are usually less than a
pixel but can be as high as 1.8 pixels for one orbit in
SBC/F140LP. For the HRC/F220W images we found that it is sufficient to 
use the pipeline combined and drizzled image that is created for each
orbit. Note that the combined and geometrically corrected ACS/SBC and
HRC images produced by {\tt multidrizzle} have a pixel scale of
$0.025\arcsec / {\rm pixel}$ and are normalized to 1 s
exposure time.
The total exposure times are 24800 s for the FUV and 
2320 s for the NUV master image and have to be considered when
calculating the photometric errors. The combined and geometrically
corrected FUV (SBC/F140LP) and NUV 
(HRC/F220W) master images are shown in Fig.~\ref{image1} and
\ref{image2}, respectively. As expected, the FUV image is considerably less
crowded than the NUV image. Both images show significant
concentrations of sources towards the cluster core. 

\subsection{Source detection}
\label{detection}

As one of our main goals of this study is to construct a deep FUV--NUV
CMD, we took special care to identify as many of the weak sources as
possible. We first used {\tt daofind} (Stetson 1991) running under
{\tt IRAF} to create an initial list of source detections. After some
experimentation, we used an FWHM of 5 pixels and a rather high
threshold to maximize the number of faint sources detected and
minimize the number of multiple detections in and around bright
sources. 
We caution that the FWHM used in {\tt daofind} does not reflect the ``true''
FWHM of the SBC point-spread-function (PSF), but is rather a
compromise to minimize the number 
of false detections at the image rims and in the defect or shaded areas
of the MAMA detectors (due to the broken anode of the ACS MAMA; see
the ACS handbook for details). We overplotted the source coordinates 
on the FUV master image and inspected them by eye. Overall, the
process worked well. However, we found that the automatic process left
some faint sources missing and others (bright sources) that were
multiply detected. We inserted missing faint sources by hand and
deleted false detections at the rims of the image, in the defect area,
and multiple detections around bright stars. In the end, our resulting
catalog of FUV sources contained 2731 entries. 

Source detection in the NUV master image worked well with standard
values for the threshold (Stetson 1991), and 9164 sources were
detected. 
In order to match the FUV and NUV catalogs, we first transformed
the FUV coordinates onto the NUV frame using the {\tt geomap} and
{\tt geoxytran} tasks running under {\tt IRAF}. We used 30 bright stars
that are common in both the FUV and NUV image as a reference list
required by {\tt geomap} to determine the geometrical transformation
between the two frames. 
Small shifts between the transformed FUV and the NUV coordinate of
sources common in both frames are to be expected. These shifts depend
on how well the applied coordinate transformation actually worked, and
also on the width of the stars' PSF in the images. We thus checked for
common pairs within a tolerance of 1.5 pixels, which resulted in 1686
matches. Increasing the tolerance to 2.5 pixels added another 193
matches. These additional matches are uniformly distributed across the
NUV master image with no preferred location, suggesting that our
coordinate transformation worked very well and no distortions are
visible. A total of 2137 FUV sources are located within the NUV field of
view. Out of these, 258 objects were not matched with an NUV
counterpart. Inspecting these unmatched sources by eye, we added
further 34 NUV sources to our initial NUV source
list. Photometry on these 34 additional NUV sources was then
performed without a Gaussian recentering as these additional sources
are very faint and a recentering algorithm will actually centre on the
next brighter star, see Sect.~\ref{photometry} and
\ref{variability}. Counterparts to the remaining 224 unmatched FUV
sources are too faint to show up in the NUV master image.    

We expect that a few false matches are among our FUV--NUV
pairs. The number of false matches depends on the number of objects in
the two source lists, the number of found pairs, and the matching
tolerance (here we chose 2.5 pixels). Following Knigge et al. (2002), we
estimate that less than $\approx 37$ pairs (1.9\% of all found matches) are
chance coincidences due to the source density in the NUV field of view.
Most of the remaining unmatched FUV sources are located
outside or at the rim of the NUV frame. 

We chose 2.5 pixels as our maximum tolerance radius as a compromise in
order to get all common pairs, but to keep the number of false matches at
an acceptable level. As can be seen from Table~\ref{catalogue}, the
actual matching radius is in most cases much smaller, with a mean
matching radius of just 0.5 pixels. 

A catalog listing all our FUV--NUV matches is available in the
electronic edition of the Journal. For reference, Table~\ref{catalogue} lists
only 10 entries. We also include sources without NUV 
counterparts in our catalog because these are likely to
include additional WD and CV candidates.

\begin{deluxetable}{ccccccccccccccc}
\tabletypesize{\scriptsize}
\tablewidth{20cm}
\setlength{\tabcolsep}{0.1cm}
\rotate
\tablecolumns{15}
\tablecaption{Catalog of all sources in  our FUV
  field of view. The first nine columns are self explanatory. The
  tenth column denotes the matching radius between the NUV and
  transformed FUV Cartesian coordinates. The radius from the cluster
  centre (at x$\approx$900 and y$\approx$850 in our FUV master
  image, determined from star number counts in both master images) is
  given in column eleven. The optical $B, V, U$ magnitudes and the
  corresponding ID from the \citet{vandermarel} optical catalog are
  given in the last four columns. An asterisk means that there is no
  entry because there is no match with an NUV source and/or
  with an optical source. Only entries 1 -- 5 and 1000 -- 1005 are
  given. Table \ref{catalogue} is published in its entirety in the
  electronic edition of the {\it Astrophysical Journal}. A portion is
  shown here for guidance regarding its form and content. \label{catalogue}}   
\tablehead{
\colhead{id} & \colhead{alpha} & \colhead{delta} & \colhead{x$_{FUV}$} & \colhead{y$_{FUV}$} & \colhead{$FUV$} & \colhead{$\Delta FUV$} & \colhead{$FUV-NUV$} & \colhead{$\Delta(FUV-NUV)$} & \colhead{tolerance} & \colhead{radius} & \colhead{$V$} & \colhead{$B$} & \colhead{$U$} & \colhead{$\rm{id}_{VDM}$}\\
\colhead{} & \colhead{[$^{\rm h} ~^{\rm m} ~^{\rm s}$]} & \colhead{[$^{\circ} ~\arcmin ~\arcsec$]} &\colhead{[pixel]}  &\colhead{[pixel]}  &\colhead{[mag]} & \colhead{[mag]} & \colhead{[mag]} & \colhead{[mag]} & \colhead{[pixel]} & \colhead{[$\arcsec$]} & \colhead{[mag]} & \colhead{[mag]} & \colhead{[mag]} &\colhead{}}
\startdata
 1 & 21 29 59.519 & 12 9 51.352 & 164.820 & 516.783 & 15.892 & 0.003 & * & * & * & 20.179 &      * &      * &      * &     *\\
 2 & 21 29 59.514 & 12 9 53.303 & 167.642 & 594.829 & 24.823 & 0.116 & * & * & * & 19.388 &      * &      * &      * &     *\\
 3 & 21 29 59.507 & 12 9 53.013 & 171.709 & 583.227 & 25.584 & 0.178 & * & * & * & 19.390 & 20.125 & 20.674 & 20.576 & 25120\\
 4 & 21 29 59.503 & 12 9 53.360 & 174.173 & 597.098 & 23.667 & 0.058 & * & * & * & 19.216 &      * &      * &      * &     *\\
 5 & 21 29 59.502 & 12 9 54.293 & 174.907 & 634.400 & 24.352 & 0.084 & * & * & * & 18.912 & 18.659 & 19.282 & 19.174 & 24915\\
1000 & 21 29 58.603 & 12 9 46.293  & 702.035 & 314.413 & 23.500 & 0.049 &      * &     * &     * & 14.275 &      * &      * &       * &       *\\
1001 & 21 29 58.602 & 12 9 45.692  & 702.384 & 290.392 & 23.018 & 0.039 &      * &     * &     * & 14.837 & 18.232 & 18.775 &  18.636 &    8759\\     
1002 & 21 29 58.602 & 12 9 56.222  & 702.513 & 711.564 & 23.993 & 0.072 &  3.154 & 0.080 & 0.690 &  6.029 & 19.673 & 20.072 &  20.012 &    5498\\     
1003 & 21 29 58.602 & 12 10 12.702 & 702.642 & 1370.79 & 17.520 & 0.005 & -0.204 & 0.007 & 0.072 & 13.923 &      * &      * &       * &       *\\     
1004 & 21 29 58.601 & 12 9 52.474  & 702.917 & 561.646 & 23.815 & 0.067 &      * &     * &     * &  8.732 &      * &      * &       * &       *\\     
1005 & 21 29 58.601 & 12 9 56.542  & 703.298 & 724.359 & 23.182 & 0.051 &  2.973 & 0.058 & 0.350 &  5.835 & 18.920 & 19.424 &  19.271 &    5396\\     
\enddata
\end{deluxetable}

\subsection{Aperture photometry}
\label{photometry}

Photometry was carried out on the combined and geometrically corrected
images for each filter using {\tt daophot} (Stetson 1991) running
under {\tt IRAF}. We used apertures of 3 pixels for  both the
FUV and the NUV data and allowed also for a Gaussian recentering
of the input coordinates derived in {\tt daofind}. In the FUV,
aperture corrections were 
determined via curves of growth constructed from isolated stars in our
master images. For the NUV data, we used the encircled energy
fractions published by Sirianni et al.~(2005). Because of close
neighbour stars in the centre of M\,15 (especially important in the
HRC/F220W images; see Fig.~\ref{image2}), we chose a small sky annulus
of 5 -- 7 pixels to minimize the number of stars within the sky annulus.
However, our small sky annulus also contains flux from the target
star. Thus, too much flux will be subtracted during sky
subtraction. For a few isolated stars, we were able to choose a larger
sky annulus of 50 -- 60 pixels. Assuming that we get the ``true'' flux
for the source we want to measure if we use the larger sky annulus, 
we can then apply a flux correction for the small sky annulus to all
sources. These flux corrections depend on the PSF of the star, which is
different for the SBC/F140LP and HRC/F220W. For the sake of
simplicity, we assume the corrections to be the same for all stars in the
master image in each filter. Table \ref{corr} lists all encircled
energy fractions and sky corrections for apertures of 1 -- 10 pixels
for both the SBC/F140LP and HRC/F220W filters.  

In order to convert count rates into fluxes and STMAGs, we used
conversion factors of 
\nonumber
\begin{eqnarray}
\nonumber
\rm{PHOTFLAM}_{SBC/F140LP} = 2.7182\times10^{-17} \rm{erg\,cm}^{2} \AA^{-1} \rm{counts}^{-1}\\
\nonumber
\rm{PHOTFLAM}_{HRC/F220W} = 7.9999\times10^{-18} \rm{erg\,cm}^{2} \AA^{-1} \rm{counts}^{-1}
\end{eqnarray}
Fluxes are then converted into STMAGs using
\begin{eqnarray}
\nonumber
\rm{STMAG} = -2.5 \times log_{10} (flux/(ee \times skycorr)) + 21.1,
\end{eqnarray}
where $\rm{flux} = \rm{count rate} \times \rm{PHOTFLAM}$ (i.e.\ the
measured flux inside our chosen aperture), ee is the encircled energy
fraction and skycorr is the flux correction for the small sky annulus
listed in Table~\ref{corr}. The conversion factors PHOTFLAM are given
in the image headers; see also the ACS data handbook.   

\begin{table}
\begin{center}
\caption{\label{corr} Encircled energy fractions and sky corrections
  for apertures of 1 to 10 pixels. The encircled energy fractions for
  the SBC/F140LP data were determined using curves of growth for
  isolated stars in the SBC/F140LP master image; the encircled energy
  fractions for the HRC/F220W were taken from Sirianni et al.~(2005).} 
\begin{tabular}{cccccc}
\tableline
\multicolumn{2}{c}{aperture} & \multicolumn{2}{c}{encircled energy
  fraction}
&  \multicolumn{2}{c}{sky correction} \\
{[pixels]} & [$\arcsec$] & HRC/F220W & SBC/F140LP & HRC/F220W & SBC/F140LP \\
\tableline
1    & 0.025       & 0.235     & 0.10       &    1.014  &  1.02 \\
2    & 0.050       & 0.506     & 0.29       &    1.014  &  1.03 \\
3    & 0.075       & 0.623     & 0.45       &    1.025  &  1.04 \\
4    & 0.100       & 0.686     & 0.54       &    1.039  &  1.06 \\
5    & 0.125       & 0.733     & 0.60       &    1.057  &  1.09 \\
6    & 0.150       & 0.758     & 0.64       &    1.081  &  1.13 \\
7    & 0.175       & 0.771     & 0.66       &    1.110  &  1.17 \\
8    & 0.200       & 0.782     & 0.68       &    1.145  &  1.23 \\
9    & 0.225       & 0.791     & 0.70       &    1.186  &  1.29 \\
10   & 0.250       & 0.801     & 0.72       &    1.235  &  1.37 \\
\tableline
\end{tabular}
\end{center}
\end{table}

\section{The FUV--NUV CMD}
\label{cmd}

\begin{figure}[h]
\centerline{
\includegraphics[width=16cm]{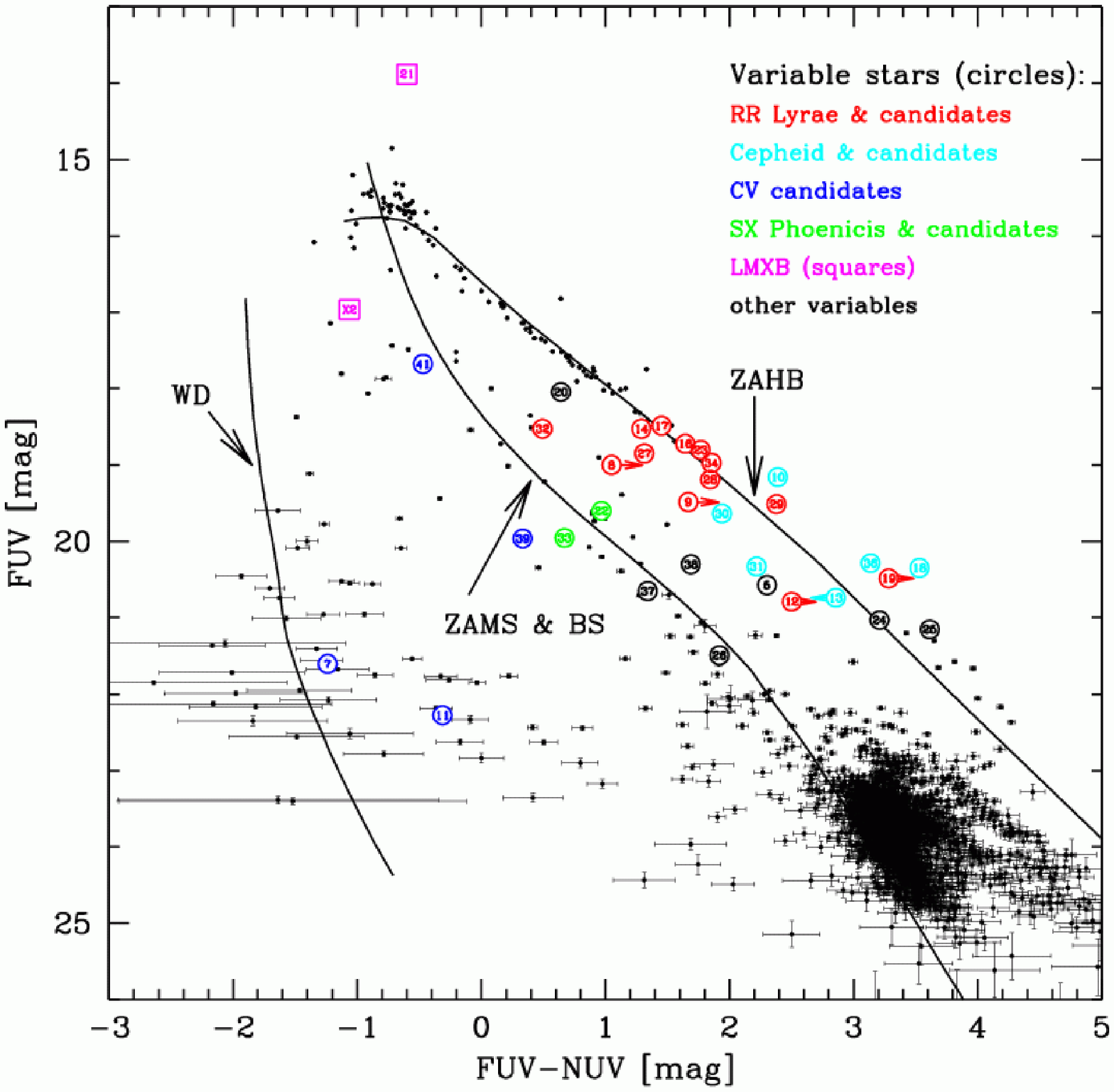}
}
\caption{\label{cmdps} FUV--NUV CMD of the core region of M\,15. For
  orientation purposes, we include a theoretical WD cooling sequence,
  a ZAMS, and a ZAHB (see
  text for details). The circles and the corresponding numbers denote
  variable FUV sources, as discussed in Sect.~\ref{variability}. The
  location of the UCXB M\,15 X-2 is also indicated with a magenta
  square. The brightest FUV star detected, V21, corresponds to the
  luminous LMXB AC\,211, see Sect.~\ref{lmxb}.}      
\end{figure}

The FUV--NUV CMD of the core region of M\,15 is shown in
Fig.~\ref{cmdps}. Several stellar populations show up, such as the
bright HB stars, BSs, and WD candidates. A number of sources are located
between the WD cooling sequence and the MS. These sources include both
detached and interacting WD-MS binaries, and we refer to these sources
as ``gap objects'', see Sect.~\ref{cv}. The region of data
points in the lower right (i.e.\ faint and red) part of the CMD
denotes the MS stars and RGs. The turnoff is at
FUV $\approx 23.5$ mag and FUV--NUV $\approx 3$ mag at the bright
end of this data region. As can be seen,
our CMD reaches at least 2 mag below the FUV MS
turnoff. A trail of stars extends from the MS turnoff
towards red and fainter magnitudes. Sources that match with optical
RGs are all located on this trail. Error bars for each data
point are also plotted.   

For orientation purposes, we have also calculated and plotted a set of
theoretical tracks representing the zero-age main-sequence (ZAMS), the
zero-age horizontal branch (ZAHB), and the WD cooling sequence. The
sequences were constructed using theoretical ZAHB models from
\citet{dorman}, the fitting formulae of \citet{tout}, 
and the \citet{wood} grid of theoretical WD cooling curves for our WD
cooling sequence. We then used {\sc synphot} within {\sc STSDAS} to
calculate the FUV and NUV magnitudes of stars on the corresponding
sequence. This was achieved by interpolating on the Kurucz grid of
model stellar atmospheres and folding the resulting synthetic spectra
through the response of the appropriate filter+detector
combinations. For the WD cooling sequence, we used a grid of synthetic 
DA WD spectra kindly provided by Boris G\"{a}nsicke (see G\"{a}nsicke
et al.\ 1995). The procedure is described in more
detail in \citet{dieball1}. For all our synthetic tracks we adopted a
distance of 10.3 kpc, a reddening of $E_{B-V}=0.1$, and a
cluster metallicity of $[Fe/H] = -2.26$ \citep{harris}. 

In the following, we describe the stellar populations within our CMD
in more detail.

\subsection{HB stars}
\label{hb}

M\,15 is one of the most metal-poor GCs in our Galaxy. Like other
metal-poor GCs, its optical CMD reveals a blue HB and a prominent blue
tail extending to fainter magnitudes (Van der Marel et al.~2002; see
also Fig.~\ref{cmd_opt}). This tail is the so-called EHB, which is
separated from the horizontal part by 
a gap. \citet{brown} suggested that the subluminous EHB stars in
NGC\,2808 (as well as the EHB gap in its optical CMD) can be explained
by a late helium-core flash that these stars undergo while they
descend the WD sequence. In NGC\,2808's FUV--NUV CMD, the HB stars
cluster in two distinct data clumps around FUV--NUV $\approx -0.7$ mag
(BHB) and FUV--NUV $\approx -1.2$ mag (EHB). 

Our FUV--NUV CMD of M\,15 reveals a strong population of 133 HB
stars along a well-defined HB sequence and within a bright data region
at FUV $\approx 15.5$ mag and FUV--NUV $\approx -0.6$ mag. This is
different from the FUV--NUV CMD of NGC\,2808 \citep{brown, dieball1},
which shows two distinct clumps of HB stars, corresponding to the
optical EHB and BHB. However, our HB stars within the bright data clump
match with optical EHB stars, whereas the sources along the ZAHB match
with BHB stars, see Sect.~\ref{optical} and Fig.~\ref{cmd_opt}
below. Thus, we possibly see a similar population of EHB stars here as
in NGC\,2808.   

\subsection{Blue Stragglers}
\label{bs}

BSs are thought to be the end product of a collision or coalescence of
two or more MS stars; i.e., they are dynamically formed objects and as
such are expected to be preferentially found in the dense cluster
cores. As they are more massive than ordinary cluster MS stars, we
expect them to be slightly evolved. In optical CMDs, they are
therefore located above and slightly to the red of the MS. In our
FUV--NUV CMD, see Fig.~\ref{cmdps}, we see a trail of stars starting
from the clump of MS stars and RGs in the lower right corner 
and reaching along the ZAMS up towards the clump of EHB stars at FUV
$\approx 15.5$ mag. This trail of stars is below the ZAHB and less clearly
defined than the sequence formed by the HB stars. The location of
these sources roughly agrees with the expected location of BSs, i.e.\
along the ZAMS but brighter than the MS turnoff. We find 69 BS
candidates but caution that a small number of these may not be BSs
since the various stellar populations in our CMD partly overlap. The
discrimination between the gap objects and BS candidates is particularly
difficult and would require further observations.

\subsection{White Dwarfs}
\label{wd}

A population of about 28 sources can be seen in Fig.~\ref{cmdps} that
lie along or close to the theoretical WD cooling curve and are
therefore probably hot, young WDs. Twenty-five of our WD candidates are
brighter than 22.5 mag.  
 
Following \citet{knigge1}, we can estimate the number of expected WDs
by scaling from the number of HB stars in the same field of view. We
count $\approx 133$\ BHB and EHB stars in our CMD. Given that the
lifetime of stars on the HB is approximately $\tau_{HB} \simeq
10^{8}$~yr (e.g. Dorman 1992), we can predict the number of WDs above
a given temperature on the cooling curve from the relation

$\frac{N_{WD}}{N_{HB}} \sim \frac{\tau_{WD}}{\tau_{HB}}$ 

where $\tau_{WD}$ is the WD cooling age at that temperature (e.g. Knigge
et al.~2002). WDs at FUV $\sim 22.5$ mag on our WD sequence have 
a temperature of $T_{eff} = 26,000$ K and a corresponding cooling age
of $\tau_{WD} \simeq 3.99 \times 10^{7}$~yr. We therefore predict
a population of approximately 53 WDs brighter than FUV = 22.5, which
strongly suggests that most if not all of our candidates are indeed WDs. 
Note that more and possibly fainter WDs are present in our FUV field
of view but are not detected in the NUV and thus do not appear in our
FUV--NUV CMD.     

We caution that this is a simplistic approach and we do not take
the effects of mass segregation into account. Richer et al.\ (1997)
suggested that if the WD cooling times are an appreciable fraction of
the cluster age (or rather the relaxation time), the slope of the
cluster initial mass function and the relation between stellar mass 
and MS lifetime become important. This can increase the expected
number of faint WDs up to a factor of 3.   

\subsection{CV candidates}
\label{cv}

As can be seen from our Fig.~\ref{cmd}, there are a number of sources
that are located between the WD cooling sequence and the ZAMS. This is
a region in the CMD where we expect to find CVs, but also detached
WD-MS binaries. As we cannot discriminate between true CVs and
detached WD-MS binaries based on our FUV--NUV CMD alone, we refer to
these sources as ``gap objects'', thus denoting sources that are
located in the gap region between the WD cooling sequence and the
MS. We find $\approx 57$ sources in this gap zone but caution again
that this figure is not to be taken as a strict number as the
discrimination especially between these gap objects and other
populations like WDs, MS stars and BSs is somewhat difficult. In the
following, we compare this number with theoretical predictions
for CVs. For this purpose, we scale the predicted number of
dynamically-formed CVs in 47\,Tuc to M\,15.  

For 47\,Tuc, \citet{distefano} predicted a population of $\sim 190$
active CVs formed via tidal capture. \citet{ivanova} investigated all
sorts of dynamical and primordial formation channels and predict a few
hundred CVs that should be present in the cluster core in 47\,Tuc. However,
\citet{shara06} found that dynamically produced CVs have shorter lives
than their field counterparts, which decreases the expected number of
CVs by a factor of 3. They based their study on $N$-body simulations
starting with 100,000 objects and a binary fraction of 5\% and found
that exchange interactions are more likely to make CVs with higher
mass MS donors (donor mass $> 0.7 M_{\odot}$), which are more likely
to be short-lived than their field counterparts. 

We follow the simplified estimate for the capture rate in GC cores
\citep{heinke},   

$\Gamma \propto \rho_{c}^{1.5} \cdot r_{c}^2 $, 

where $\rho_{c}$ is the central luminosity density and $r_{c}$ is the
core radius. Taking $\rho_{c}$, $r_{c}$, and the distance to the clusters 
from \citet{harris}, we find that the expected number of dynamically-formed
CVs in M\,15 is $\approx 200$, comparable to what is expected for
47\,Tuc. \citet{distefano} found that approximately half of the
captures would take place within 1 core radius. Given our detection
limits, we cannot hope to detect the very oldest and faintest CVs.  
The coolest WDs that we detect have $T_{eff} = 26,000$ K (see
Sect.~\ref{wd}). This suggests that we can only expect to find
relatively bright, long-period CVs above the period gap (see Townsley
\& Bildsten, 2003, their Figs.~1 and 2). About 20 of these
long-period CVs should exist in 47\,Tuc (Di\,Stefano \& Rappaport,
1994, their Fig.~3 and Table~5). 
Scaling this number, we expect 21 such systems in M\,15. As our image
covers more than $5 \times r_{core}$, we expect to find most, but
perhaps not all, of these sources. As a rough estimate, we suggest that
10 -- 21 of the long-period, active CVs should be within our field of view.   
If dynamical effects shorten the lifetime of CVs by a factor of about
3, as suggested by \citet{shara06}, the expected number of CVs in our
data would drop accordingly. We caution that these are extremely rough
estimates, but they suggest that some, but not all, of our gap objects
are likely to be CVs. 

\subsection{Optical counterparts}
\label{optical}

An optical catalog of 31,983 sources in the central region of M\,15
was presented by \citet{vandermarel}. These data were obtained in
1994 April in the context of {\it HST} program GO-5324, using the
WFPC2/F336W, F439W, and F555W camera/filter combination, with the PC
centred on the cluster centre. The chosen WFPC2 filters correspond
roughly to the Johnson $U$, $B$, and $V$ bands. First results from these
observations were reported by \citet{guhathakurta}. \citet{vandermarel} 
applied an improved astrometric and photometric calibration to the
original star list, and we used this catalog to find optical
counterparts for our FUV sources. 

We first transformed the $\alpha, \delta$ coordinates in the optical
catalog to the Cartesian ($x$, $y$) FUV coordinate system, using the
{\tt IRAF} task {\tt wcsctran}. Note that since the optical $\alpha,
\delta$ coordinates do not correspond exactly to the FUV ones, the
Cartesian coordinates also do not match exactly and require a further
transformation. Thus, we selected 80 HB stars that are common to
both catalogs and could be easily identified. These 80 sources
served as our reference list for a coordinate transformation. Using
the {\tt IRAF} tasks {\tt geomap} and {\tt geoxytran}, we transformed the
physical FUV coordinate system to the optical ($x$, $y$)
one.\footnote{We found a rather large shift between our FUV and the 
  optical coordinates given in the electronic version of the optical
  catalog available via CDS, Strasbourg. For example, the
  coordinates for the well-known  LMXB AC\,211 (optical id 5435) are
  $21^{h} 29^{m} 57.91^{s}  +12^{\circ} 10\arcmin 06\farcs8$ in the
  VizieR version of the optical catalog. In our catalog presented in
  Table~\ref{catalogue}, AC\,211 (FUV ID 1580) is at $21^{h} 29^{m}
  58.238^{s} +12^{\circ} 10\arcmin 1\farcs65$, which is much closer to
  previously published figures \citep{wa, kulkarni}. However, the
  original electronic table in \citet{vandermarel} listed only $\Delta
  \alpha$ and $\Delta \delta$ with respect to cluster centre and no
  absolute $\alpha, \delta$. Applying $\Delta \alpha$ and $\Delta
  \delta$ to the cluster centre coordinates given in
  \citet{vandermarel} results in $21^{h} 29^{m} 57.965^{s}
  +12^{\circ} 10\arcmin 02\farcs85$ for AC\,211, closer to previous
  published values and different from the coordinates computed by
  VizieR.}  
Allowing for a matching radius of 1.5 pixel, we then found 1632 sources
that are common to both the optical and our FUV catalog. 

\begin{figure}
\centerline{
\includegraphics[width=16cm]{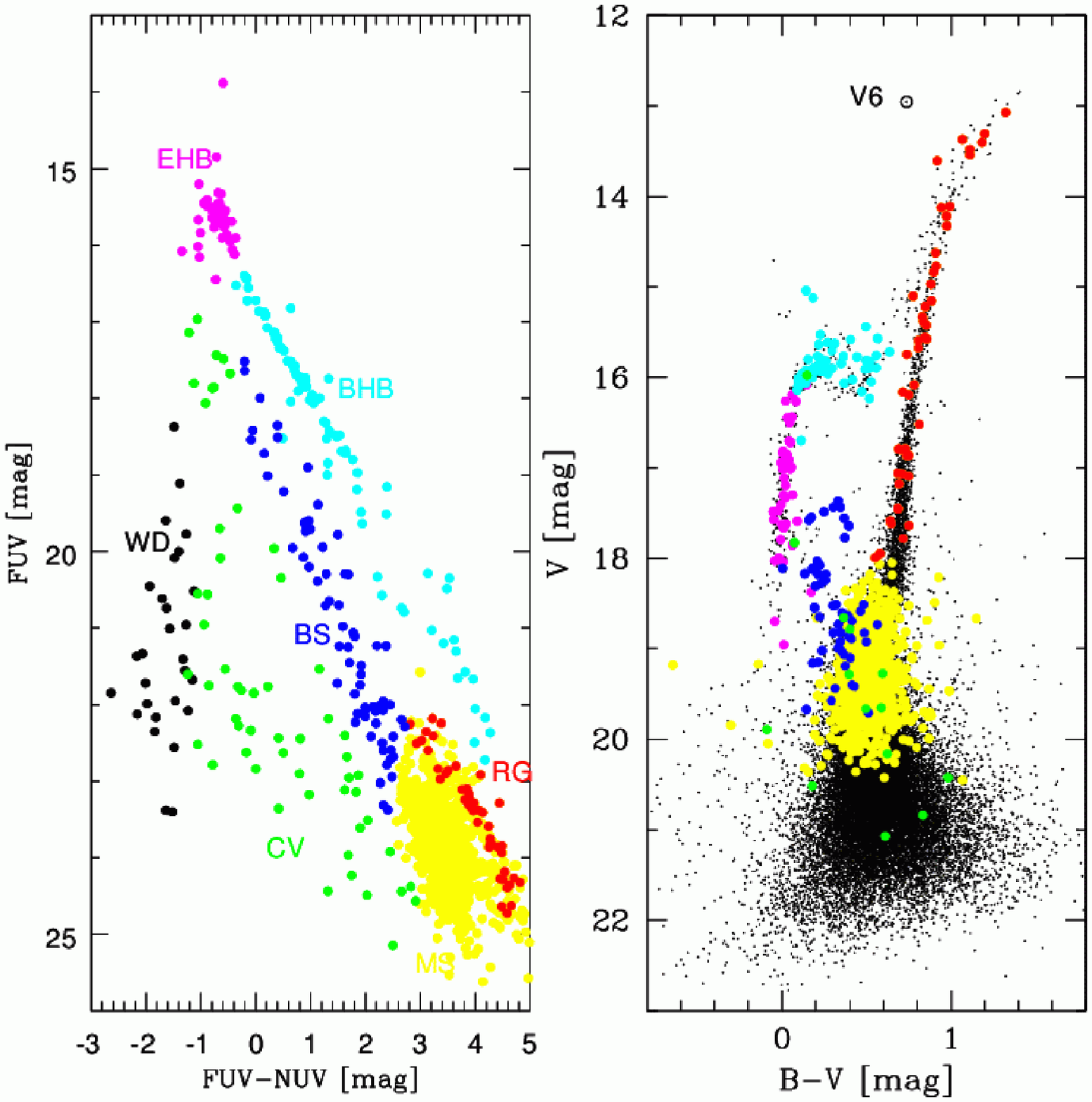}
}
\caption{\label{cmd_opt} Comparison between the various stellar
  populations found in our FUV--NUV CMD (left) and the optical CMD
  (right; based on the \citet{vandermarel} catalog). The stellar
  populations are at their expected location in the optical CMD;
  i.e., our BS candidates are mainly above the MS turnoff in the
  optical CMD, etc. (see text for details). Gap objects (which
  include CV candidates) are plotted in green, BS candidates in blue,
  MS stars in yellow, RGs in red, EHB stars in magenta and BHB stars
  in cyan. Note that we did not find any optical counterparts to our
  WD candidates. The counterpart to our variable candidate V6 is
  indicated in the optical CMD (right).}       
\end{figure}

We checked the location of the common sources in both the FUV--NUV
and the optical CMD and found that indeed all stellar populations are
at their expected location in the optical CMD, i.e.\ our matched HB
stars are on the optical blue or extended blue horizontal branch (see
Sect.~\ref{hb}), and the matched sources within our BS region
are in the BS region in the optical CMD, and so on. This is
illustrated in Fig.~\ref{cmd_opt}, where we indicate the different
stellar populations found in our FUV--NUV CMD with different colours
and overplotted them on the optical CMD. A few BSs are within the
optical MS region, which indicates that these BSs are 
formed from the coalescence of lower mass MS stars and thus are
of a lower
mass than the BS above the optical MS turnoff. All matched sources
that are located in the MS/RG clump in our FUV--NUV CMD correspond to
optical MS stars or RGs. We did not find any
counterparts to our WD candidates, which is to be expected as WDs
are extremely faint in the optical. We found 12 matches among
our gap objects. Ten of these are located along the optical
MS. These systems might be CVs or detached WD-MS binaries that are
dominated by the MS star in the optical, whereas the systems are
dominated by the accretion disk or a hot WD in the FUV. Note that
the gap object with the bluest optical counterpart is actually the
UCXB M\,15 X-2.  

\subsection{Radial distributions of the stellar populations}
\label{cumulativetxt}

\begin{figure}[ht!]
\centerline{
\includegraphics[width=15cm]{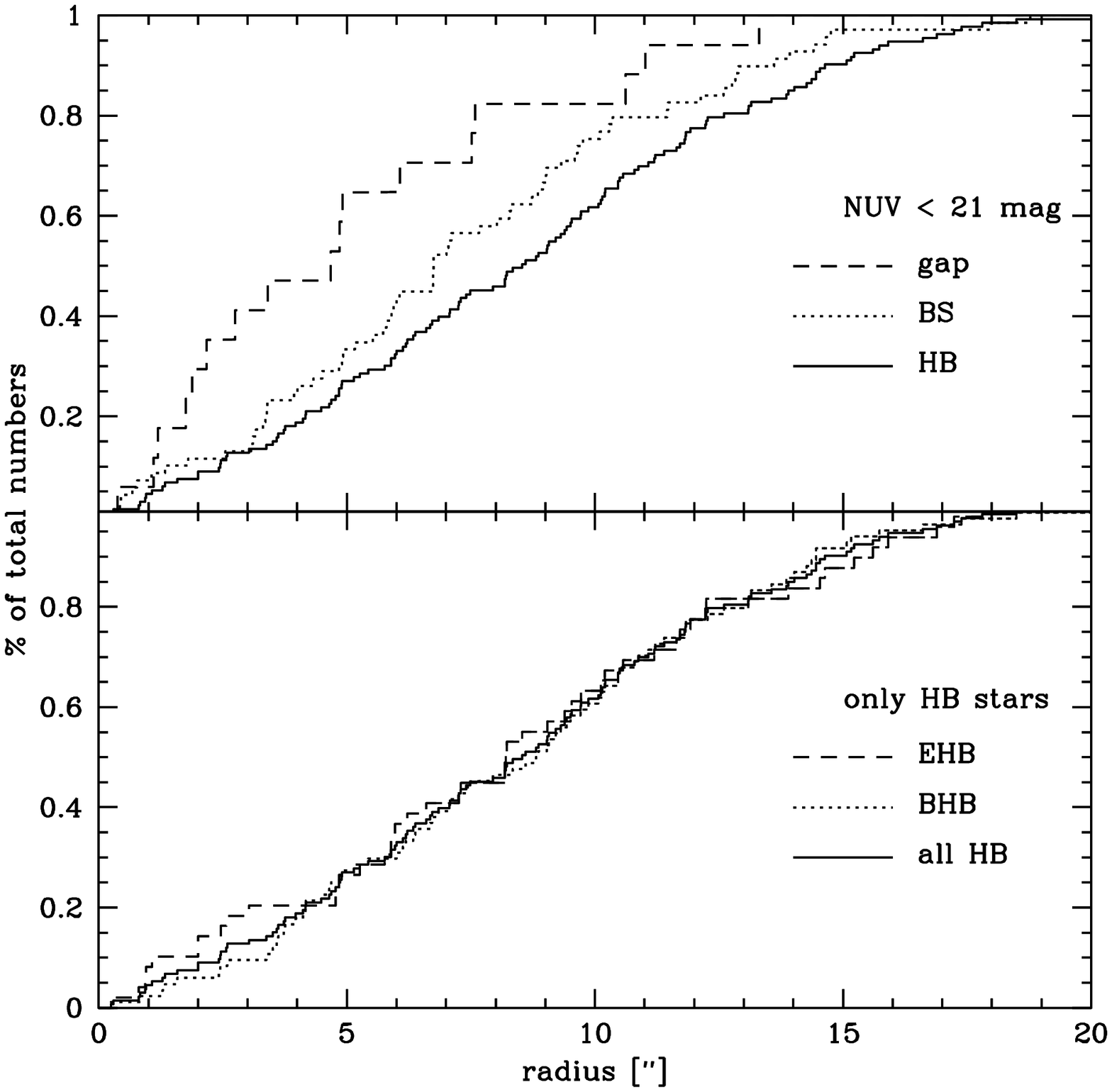}
}
\caption{\label{cumulative1} Cumulative radial distributions of the
  various stellar populations that show up in our FUV--NUV
  CMD. Top: Only sources with NUV $\le 21$ mag are
  considered. We only compare BS candidates (dotted line), gap
  objects (dashed line), and HB stars (solid line), as
  only two WD candidates are present that are brighter than 21 mag in the
  NUV. Bottom: Radial distribution of BHB stars (dotted
  line) and EHB stars (dashed line). EHB stars seem to be
  slightly more concentrated in the core. See text for details.} 
\end{figure}

\begin{figure}[ht!]
\centerline{
\includegraphics[width=15cm]{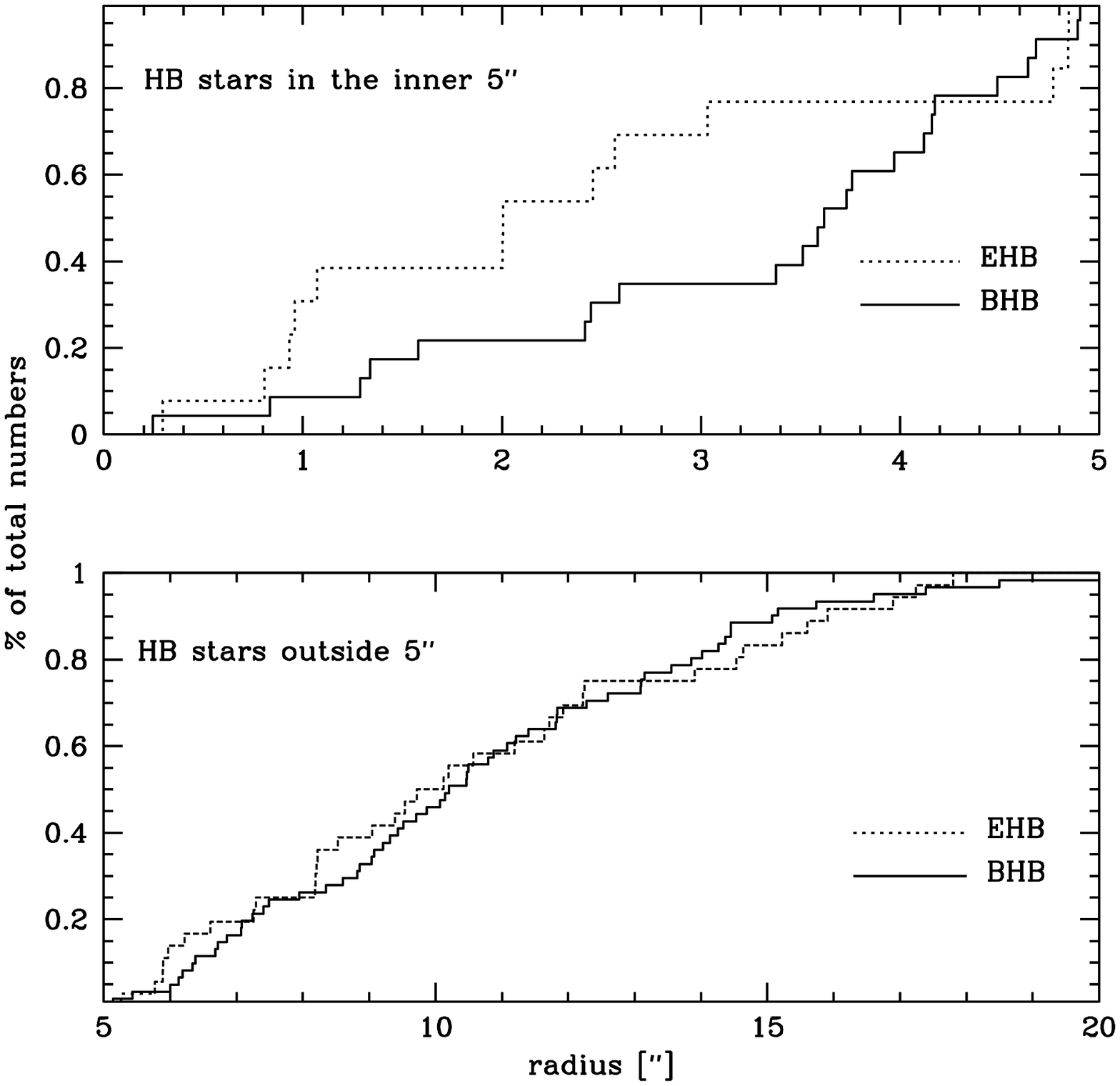}
}
\caption{\label{cumulative2} Cumulative radial distributions of EHB
  and BHB stars in the inner $5\arcsec$ (top) and outside the
  core region (bottom).}
\end{figure}

Fig.~\ref{cumulative1} shows the cumulative radial distribution of the
stellar populations that show up in our CMD and that are discussed
above. As our CMD is limited by our NUV data, we only consider data
that are brighter than NUV = 21 mag. We note that the radial
distribution of MS stars is not shown because the difference between
the MS stars and all other populations might be an effect of
incompleteness in the cluster core in the FUV image. A few bright
stars are concentrated in the cluster centre, and the broad PSF wings
of the bright sources hamper the detection of faint objects,
especially MS stars and WDs. As only two WD candidates with NUV < 21
mag are present in our limited data set, we show only the radial
distributions of BS candidates, gap objects, and HB stars, see the top
panel in Fig.~\ref{cumulative1}.  
All of the gap sources are CV candidates to start with, but we caution
again that this sample might also contain detached WD-MS binary
candidates. The gap objects seem to be the most centrally concentrated
stellar population, followed by BS candidates and then HB stars. 

We applied a Kolmogorov-Smirnov (K-S) test on the restricted
data set in order to determine the statistical significance of the
differences between the various stellar populations. However, we
caution that HB stars, gap sources, and BS candidates cannot be clearly
distinguished based on the CMD alone, and we might have sources 
in one sample that actually belong to another. Our selected
data set contains 133 HB stars, 69 BS candidates, and 16 gap sources. 

The K-S test returns the probability that the maximum difference between
the two distributions being compared should be as large as observed,
under the null hypothesis that the distributions are drawn from the
same parent populations. In this sense, a high K-S probability agrees
with both distributions to be likely from the same parent
distribution. The smaller the returned K-S probability, the more
unlikely it is that they originate from the same parent distribution,
i.e.\ the more different are the two compared distributions.
BS candidates and gap sources differ from the HB population
with K-S test probabilities of 14\% (BSs vs. HBs) and 2\% (gap sources
vs. HBs). Gap objects and BS candidates are different with K-S
probability of 8\%. 

As shown in Fig.~\ref{cumulative1}, gap sources and BS candidates are
the most centrally concentrated populations. 
There are two possible physical explanations for this: (1) Since CVs
and BSs are expected to be more massive than MS and HB stars, the
differences in radial distribution could be an indication of mass
segregation; i.e., the heavier objects sink towards the cluster core
and end up more centrally concentrated than the lighter MS and HB
stars (and WDs). (2) Alternatively, since CVs and BSs in GCs can form
through two- or three-body interactions that take 
place preferentially in the denser cluster core, they may be
overabundant there because this is their birthplace. 

In any case, the enhanced central concentration of the gap sources
provides additional evidence that most of these sources are indeed CVs 
or non-interacting WD/MS binaries (as opposed to chance
superpositions, foreground stars, etc.), i.e.\ they either segregated 
towards the cluster centre or formed in the dense core region.   

The bottom panel of Fig.~\ref{cumulative1} shows the radial
distribution of only HB stars. We compare EHB and BHB stars, and it
seems that EHB stars are slightly more concentrated in the inner $5
\arcsec$, i.e.\ the cluster core. 
As a next step, we compare the EHB and BHB stars only in the cluster
core (Fig.~\ref{cumulative2}, top panel) and outside the core region
(Fig.~\ref{cumulative2}, bottom panel). We found that the two HB
populations in the inner $5\arcsec$ are different with K-S probabilities of
7.5\%, whereas the EHB and BHB populations in the outskirts (i.e.\
outside the cluster core) of M\,15 appear not to be different, with a K-S
test probability of 93.2\%. This might indicate that EHB stars are
indeed more centrally concentrated than BHB stars. One possible
explanation could be that EHB stars have a dynamical origin
\citep{fusipecci} and could have formed from mergers of WD binary 
systems \citep{iben}. 

\section{Variable candidates}
\label{variability}

\subsection{Detection of Variability}
\label{vardetect}

For our variability study only relative magnitudes are needed. For
this purpose, we performed photometry on the individual drizzled,
i.e. geometrically corrected, images, allowing also for a Gaussian
recentering of the input coordinates. We only used exposures
taken in SBC/F140LP, since this data set provides the longest time
coverage (from 2004 October 14 to December 5). In total 90 exposures
were used for this analysis. Eighty of these had exposure times of 300
s, six were exposed for 40 s and four had exposure times of 140
s (see Table~\ref{obslog}).

\begin{figure}
\centerline{
\includegraphics[width=15cm]{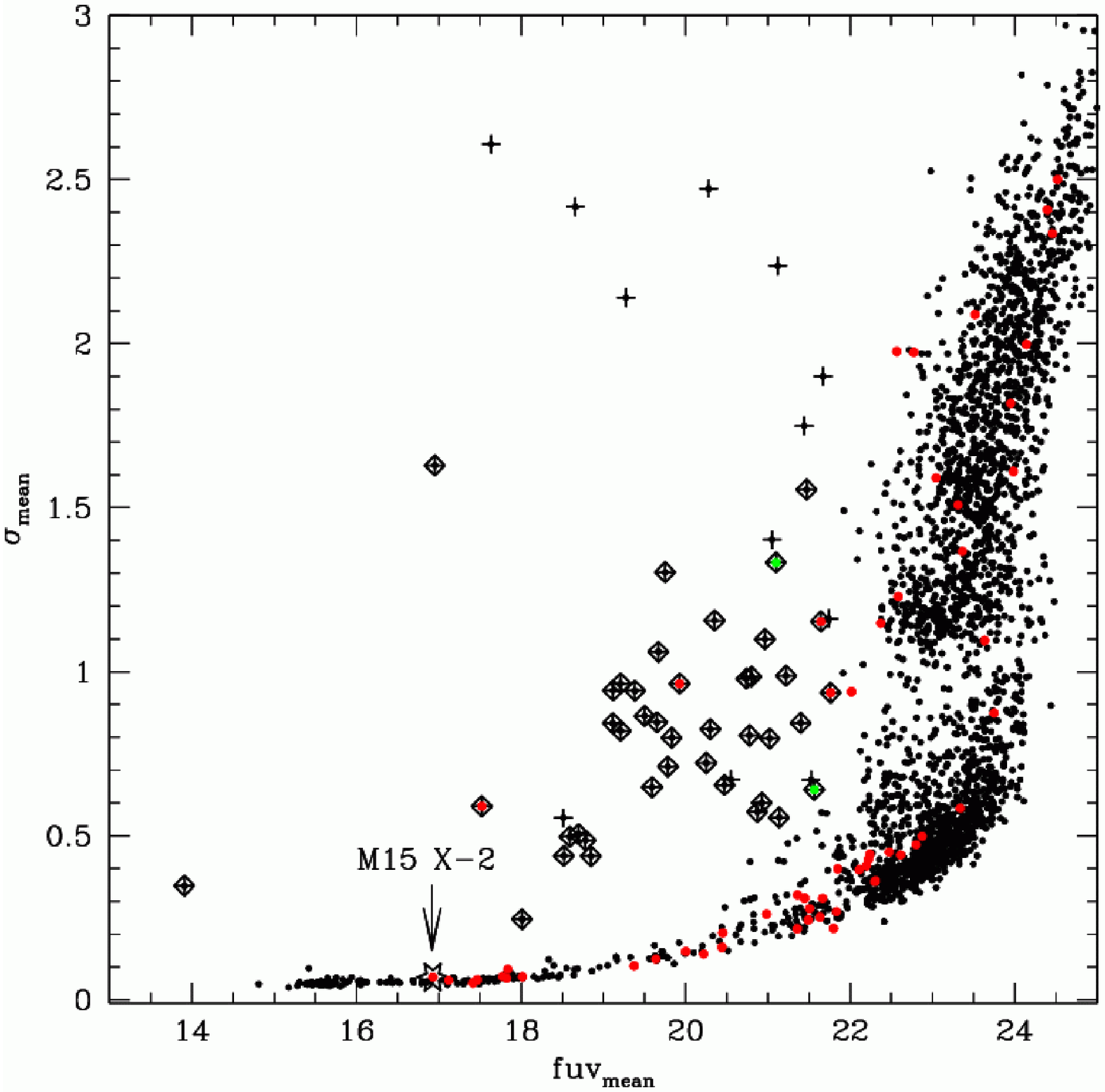}
}
\caption{\label{varsigma} Mean FUV magnitudes vs.\
  $\sigma_{mean}$ derived from the individual photometry of each
  star. The 53 initial variable candidates are marked with plus signs.
  The 41 variable candidates that remain after closer inspection on
  the individual input images are additionally marked with
  diamonds. The position of M\,15 X-2 is marked with a star and
  indicated with an arrow. The gap objects/CV candidates
  that we selected based on our FUV--NUV CMD are marked with red filled
  circles. The two additional objects (V15 and V40) that do not have an
  NUV counterpart but are likely CVs because of their variability are
  marked with green filled circles. See text for details.}     
\end{figure}

For each star in our input catalog of FUV sources, its mean magnitude
and the corresponding standard derivation were calculated from all
single magnitudes. Fig.~\ref{varsigma} shows a plot of all mean
magnitudes versus their corresponding $\sigma_{mean}$ for all stars. 
Based on this plot, we selected 53 sources that show a
$\sigma_{mean}$ exceedingly higher than their companions with similar
brightness. These initial variable candidates are marked with plus signs
in Fig.~\ref{varsigma}. As a next step, we checked each of our initial
variable candidates by eye on all 90 input images. We found that in
a few cases, the recentering algorithm 
actually centred on the nearest neighbour instead of the source
in question. This happened especially for our variable candidate
32, which for a few orbits became extremely faint. In these cases,
we forced the photometry on the initial input coordinates without
recentering. We caution that for accurate photometry a recentering is
usually better, as small shifts between the input images might occur
(see Sect.~\ref{detection}).  
However, in the case mentioned above, the recentering fails to work
and photometry forced on the initial input coordinates is then the
only option available, but it should be considered with some care. 

A few initial variable candidates were located in the halos of close
bright stars. These stars either are only detectable in the combined
master image, but are too faint to show up in the single input images
(which can have exposure times as low as 40 s), or they might just
be false detections in the halos of bright stars. For our further
analysis, these questionable candidates are not considered. Most of
these doubtful objects have no NUV counterpart (except for source
2388), which again indicates that they are either very blue and faint
sources or indeed false detections. Two variable candidates
turned out not to be true sources at all but bad pixels and were
consequently deleted from our list of FUV sources in
Table~\ref{catalogue}. We checked that there are no further known bad
pixels that coincided with suspected sources. The remaining 41
variable candidates are marked with diamonds 
in Fig.~\ref{varsigma} and are discussed in detail in the 
following. Note that six of our good variable candidates (V1, V2,
V3, V4, V6, and V35) are located outside the NUV field of view. V15
and V40 are inside the common field of view but do not have an NUV
counterpart within our chosen tolerance radius of 2.5 
pixels. Consequently, these eight variable candidates do not appear in the
CMD in Fig.~\ref{cmdps}. Our notation ``Vxx'' refers to the variable
candidates that we discuss in the present paper. We caution that
\citet{clement01} use the same nomenclature but for different stars.

Note that the UCXB M\,15 X-2 does not show a noticeable
$\sigma_{mean}$. This source has an orbital period of $22.5806 \pm
0.0002$ minutes, and the signal is clearly significant, see
\citet{dieball2}. However, the semi-amplitude of its modulation is
only $0.062 \pm 0.004$ mag, which is below our selection criterion in
the context of this paper. All our variable candidates have
$\sigma_{mean} > 0.2$ mag, and M\,15 X-2 would not have been detected
as a variable based on Fig.~\ref{varsigma}, in contrast to the LMXB
AC\,211, which corresponds to our V21. 
This shows that we detect stars that show strong variability, but we
do not claim to be complete and detect ``all'' variable stars in the
core of M\,15. For this, a more thorough variability analysis would be
needed, which is beyond the scope of the present paper.
We marked the position of the UCXB with a star and indicate it with an
arrow in Fig.~\ref{varsigma}.   
  
In order to check for systematic trends in our data, we analyzed the
light curves of the 10 brightest, non-variable stars in our
sample. A slight trend of somewhat fainter magnitudes towards the end
of each orbit is visible in all of them. The resulting periodograms
all peak at a period that is consistent with the {\it HST's} orbital
period. 

\begin{figure}
\centerline{
\includegraphics[width=17cm]{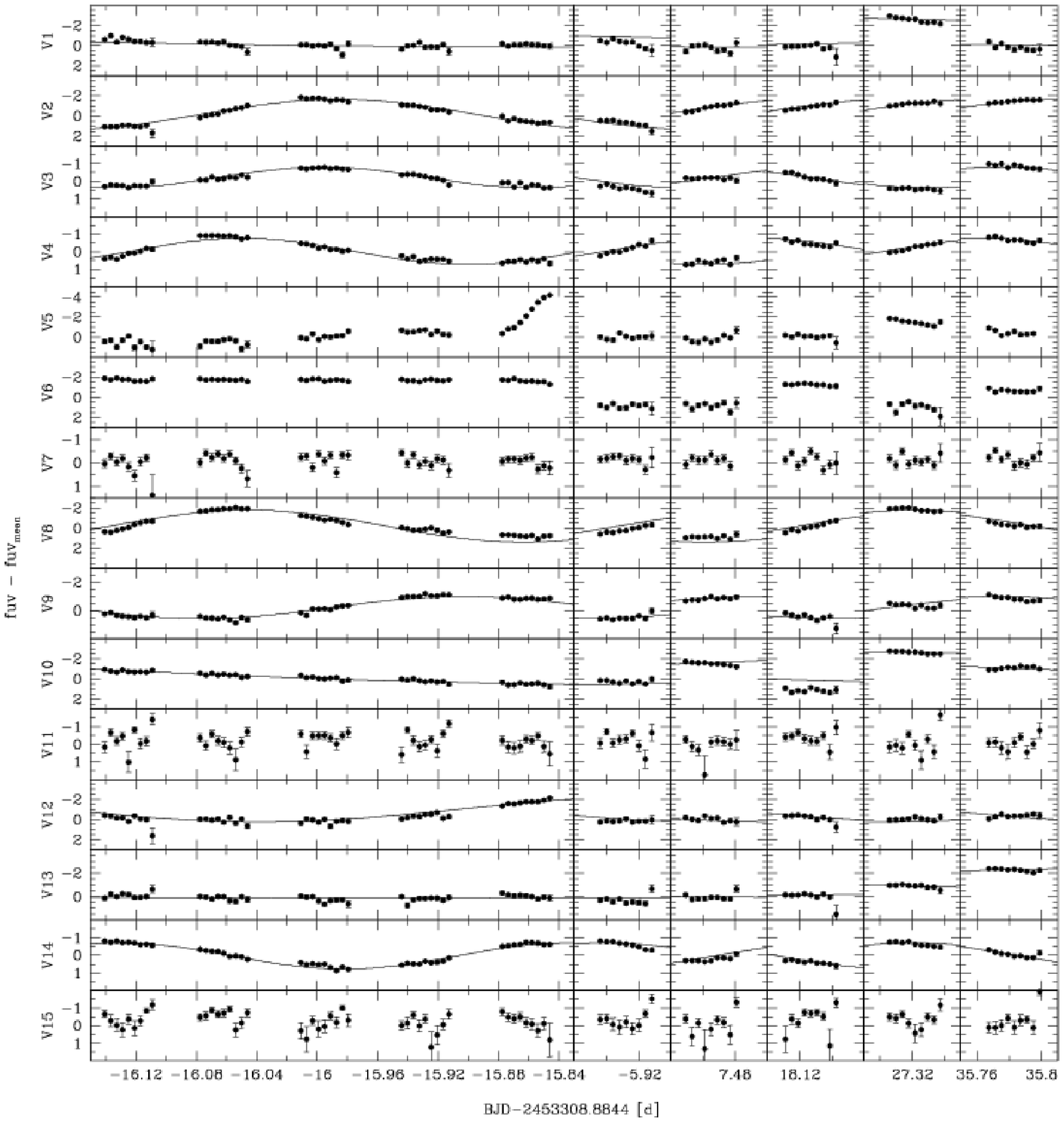}
}
\caption{\label{lcurve1} Light curves for our variable sources V1
  -- V15. The mean subtracted instrumental magnitudes are plotted
  vs.\ time.}   
\end{figure}

\begin{figure}
\centerline{
\includegraphics[width=17cm]{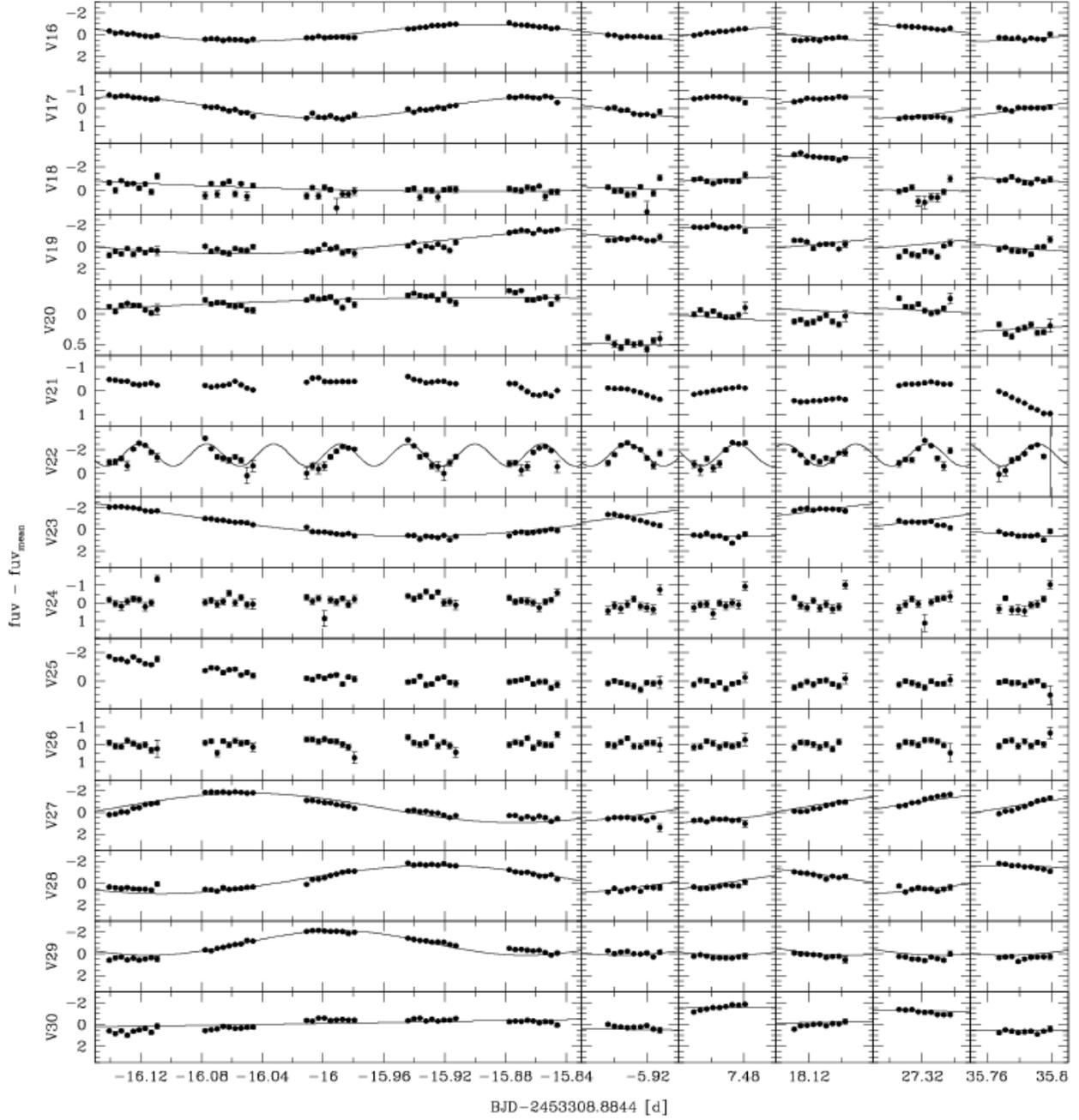}
}
\caption{\label{lcurve2} Same as Fig.~\ref{lcurve1}, but for the
  variable sources V16 -- V30.}
\end{figure}

\begin{figure}
\centerline{
\includegraphics[width=17cm]{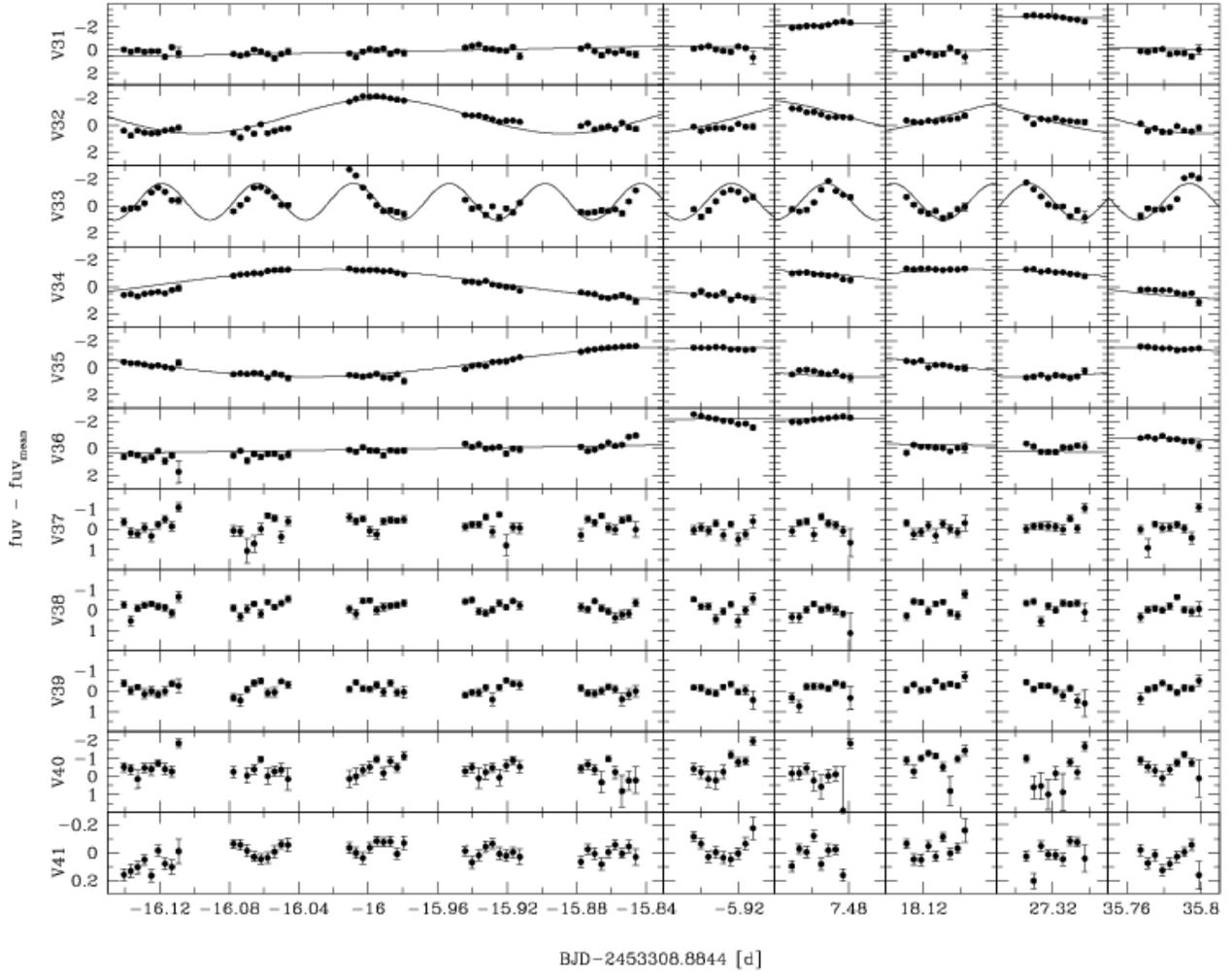}
}
\caption{\label{lcurve3} Same as Fig.~\ref{lcurve1}, but for the
  variable sources V31 -- V41.}   
\end{figure}

\begin{figure}
\centerline{
\includegraphics[width=16cm]{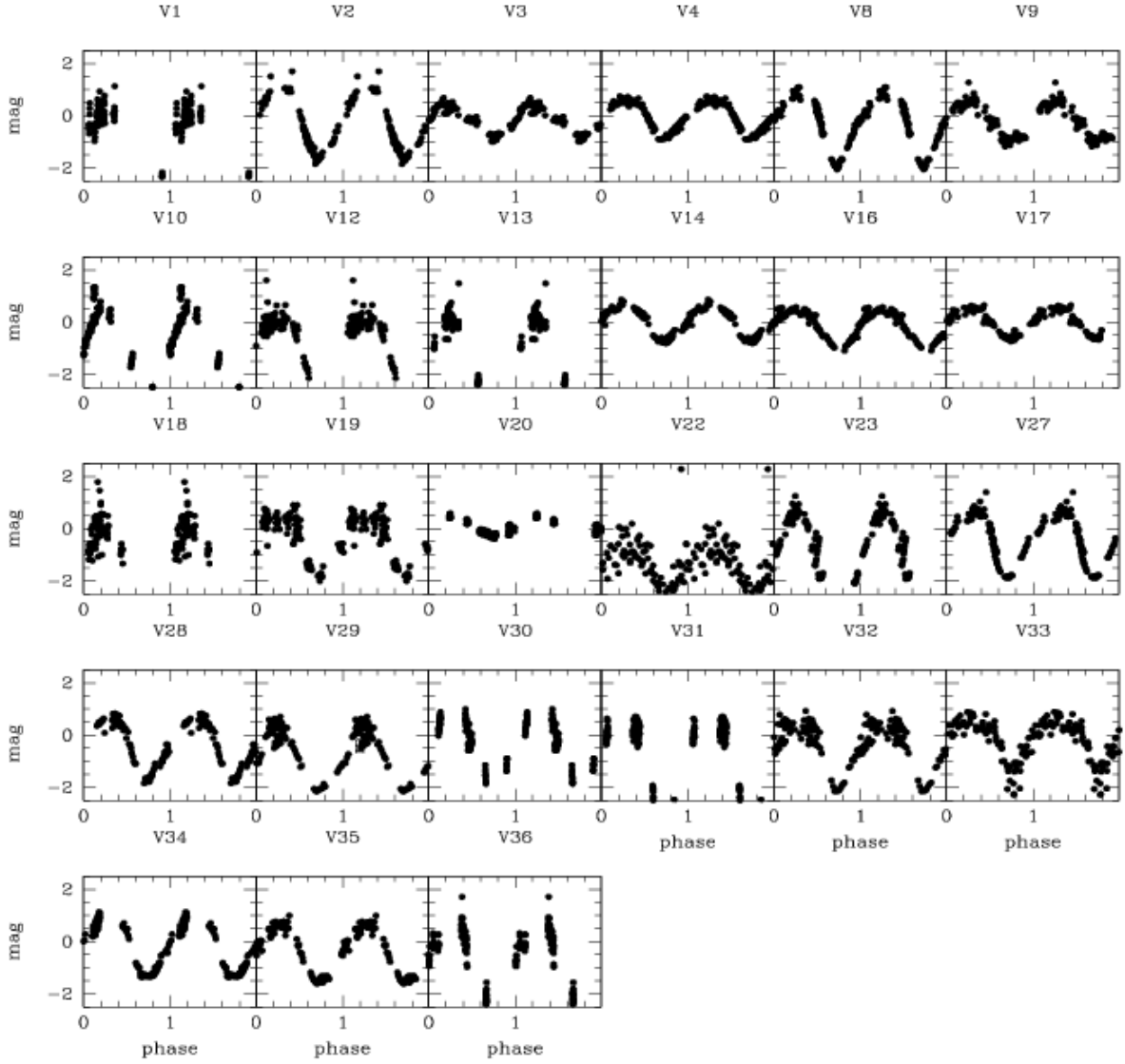}
}
\caption{\label{fold} Folded light curves for the variables
  for which we derived periods. See also Table~\ref{periods1} and
  \ref{periods2}.}     
\end{figure}

Fig.~\ref{lcurve1} to Fig.~\ref{lcurve3} show the mean-subtracted
light curves over all observing epochs for all remaining good variable
candidates. We computed $\chi^{2}$ periodograms for all data
sets in order to determine periods for all variable candidates. However, 
even if present, a periodicity can only be determined if the
observations sample that period sufficiently well. This is the case
for many but not all variables. A sine wave representing the
corresponding period is then overplotted on their light curve. The
folded light curves are shown in Fig.~\ref{fold}. Tables~\ref{periods1}
and \ref{periods2} give an overview over all variable candidates,
their periods, and their classification based on the following discussions.
     
\begin{figure}
\centerline{
\includegraphics[width=16cm]{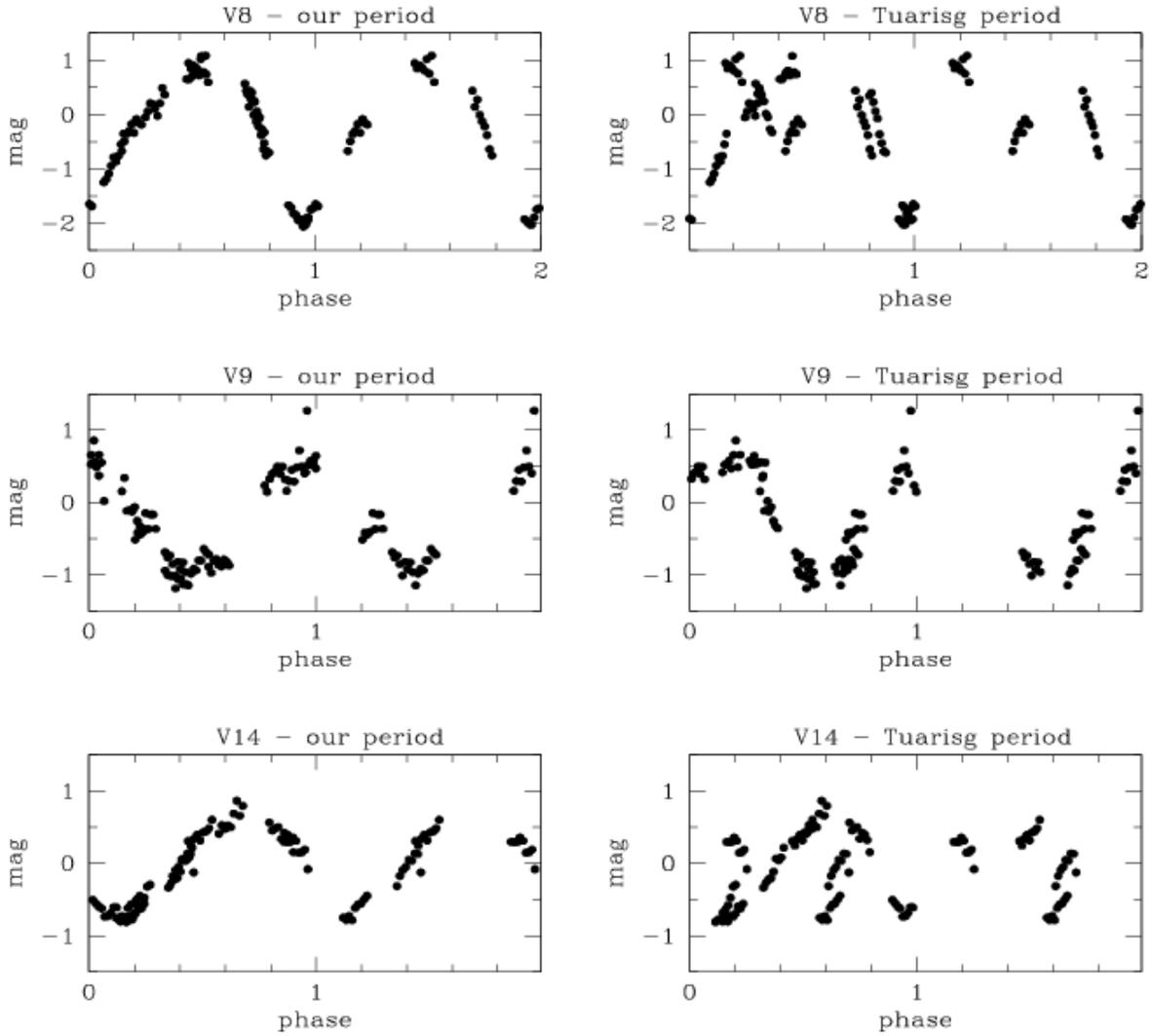}
}
\caption{\label{foldrr} Folded light curves for the RR Lyrae V8, V9, and
  V14. The left panels shows the light curves folded with the periods
  derived in this paper, while the right panels shows the light curves folded
  with a period derived by \citet{tuairisg}. See text for details.}    
\end{figure}

\begin{table}
\begin{center}
\caption{\label{periods1} Known RR Lyrae stars and RR Lyrae
  candidates. Reference for the known RR Lyrae is \citet{tuairisg}.}
\begin{small} 
\begin{tabular}{lccll}
\\
\tableline
id    & $<FUV>$  & $<FUV-NUV>$ & period & other name\\
      & [mag]  & [mag]     & [d]    &          \\ 
\hline
V2    & 18.845 & --        &  $0.3674$ & BSR\,1856, V144?\\
V3    & 18.795 & --        &  $0.2768$ & BSR\,1856, V144?\\
V4    & 18.619 & --        &  $0.2995$ & BSR\,1856, V144 (most likely counterpart)\\
V8    & 18.999 & 1.300     &  $0.3644$ & V165?\\
V9    & 19.485 & 1.917     &  $0.3492$ & V128\\
V12   & 20.790 & 2.700     &  $0.5350$ & \\
V14   & 18.526 & 1.290     &  $0.2997$ & V137?\\
V16   & 18.722 & 1.643     &  $0.3083$ & \\
V17   & 18.485 & 1.452     &  $0.2811$ & \\
V19   & 20.485 & 3.483     &  $0.5408$ & \\
V23   & 18.801 & 1.768     &  $0.6161$ & \\
V27   & 18.849 & 1.314     &  $0.3531$ & V160 \\
V28   & 19.189 & 1.845     &  $0.3615$ & V135 \\
V29   & 19.512 & 2.379     &  $0.2483$ & \\  
V32   & 18.525 & 0.493     &  $0.2110$ & \\
V34   & 18.972 & 1.854     &  $0.3973$ & \\
V35   & 19.398 & --        &  $0.4127$ & \\
\tableline
\end{tabular}
\end{small}
\end{center}
\end{table}

\begin{table}
\begin{center}
\caption{\label{periods2} Classification of all other variables.
  Periods are in days unless otherwise stated. References
  for the known variables are: (a) \citet{tuairisg}, (b) Ilovaisky et
  al.~(1993), (c) \citet{kravtsov}, (d) \citet{dieball2}.}
\begin{small} 
\begin{tabular}{lcclll}
\\
\tableline
id    & $<FUV>$ & $<FUV-NUV>$ & period & type             & other name\\
      & [mag]   & [mag]       & [d]    &                  &          \\ 
\hline
V1  & 20.967 & --        & $2.6$         & CEP candidate    & \\
V5  & 20.573 & 2.301     & --               & ?                & \\
V6  & 19.985 & --        & --               & ?                & \\
V7  & 21.604 & -1.238    & --               & CV candidate     & \\
V10 & 19.157 & 2.389     & $1.6$         & CEP              & V16 (a)\\
V11 & 22.277 & -0.313    & --               & CV candidate     & \\
V13 & 20.739 & 2.656     & $3.4$         & CEP candidate    & \\
V15 & 21.364 & --        & --               & CV candidate     & \\
V18 & 20.347 & 3.530     & $1.4$         & CEP candidate    & \\
V20 & 18.043 & 0.641     & $1.5$         & ? (BHB zone)     & \\   
V21 & 13.881 & -0.598    & $17.1$ [hrs] (b) & LMXB             & AC211 (b)\\
V22 & 19.599 & 0.968     & $1.06$ [hrs]     & SXPHE candidate  & \\
V24 & 21.030 & 3.209     & --               & ? (BHB zone)     & \\
V25 & 21.153 & 3.613     & --               & ? (BHB zone)     & \\
V26 & 21.492 & 1.919     & --               & ? (BS zone)      & \\
V30 & 19.635 & 1.938     & $3.8$         & CEP candidate    & \\
V31 & 20.328 & 2.218     & $3.8$         & CEP candidate    & \\
V33 & 19.954 & 0.670     & $1.32$ [hrs]     & SXPHE           & VGG 10041 (c)\\
V36 & 20.285 & 3.137     & $4.5$         & CEP candidate    & \\
V37 & 20.651 & 1.342     & --               & ? (BS zone)      & \\
V38 & 20.300 & 1.689     & --               & ? (BS zone)      & \\ 
V39 & 19.962 & 0.333     & --               & CV candidate     & \\
V40 & 21.692 & --        & --               & CV candidate     & \\
V41 & 17.676 & -0.469    & --               & CV candidate     & \\
--  & 16.964 & -1.062    & $22.6$ [min] (d) & UCXB            & M\,15 X-2 (d)\\
\tableline
\end{tabular}
\end{small}
\end{center}
\end{table}

\subsection{RR Lyrae stars with possible counterparts: V2, V3, V4, V8,
  V9, V14, V27, V28} 
\label{rrlyrae}

RR Lyrae stars are radially pulsating giant stars of spectral class A
-- F that change their brightness with periods of 0.2 to 1.2 days and
 amplitudes from 0.5 -- 2 mag in $V$ \citep{GCVS}\footnote{The GCVS is
  also available at: http://www.sai.msu.su/groups/cluster/gcvs/gcvs/}.  
RR Lyrae are divided into (at least) two subgroups:
RRab show asymmetric light curves and pulsate in the fundamental
tone. They show periods from 0.3 to 1.2 days and amplitudes from 0.5
to 2 mag in $V$. RRc show first-overtone pulsations, their light curves
are roughly sinusoidal with periods of 0.2 -- 0.5 days and their
amplitudes are not greater than 0.8 mag in $V$. 

The light curve of our variables V2, V3 and V4 show long-term 
variability, with periods of 0.3674, 0.2768 days, and
0.2995 days, respectively. Periods and amplitudes agree
with these variables being RR Lyrae stars. All of these three sources
match within $2\arcsec$ with the known RR Lyrae V144 in
\citet{clement01}, or BSR\,1856 in \citet{tuairisg}, and constitute
possible counterparts to this object. V2's coordinates match within
$0\farcs5$ with the known RR Lyrae and are thus the closest positional
match, whereas our V3 and V4 both match within $1\farcs8$ with BSR
1856. \citet{tuairisg} derived a period of 0.2990 days for BSR\,1856,
which agrees  very well with the period of our V4. It is likely that
V2 and V3 maybe are newly discovered RR Lyrae whereas V4 is the true
counterpart to BSR 1857.  

The coordinates of V8 match within $0\farcs9$ with V165 in
\citet{tuairisg}, for which they derived a period of 0.4339 days. We
derived a much smaller period of 0.3644 days, which fits our
observed light curve better than a period of 0.4339 days. As a further
test, we folded the light curve with our period
and a period of 0.4339 days, see Fig.~\ref{foldrr}. As can be seen, the
resulting plot with the longer period does not result in a smooth
curve. Although the coordinates agree well enough, the different
periods suggest that our V8 is probably not the same source as V165 in
\citet{tuairisg}. 

V9 matches within $1\farcs1$ with V128 in \citet{clement01}. However,
our period is 0.3492 days and thus somewhat shorter than the one
\citet{tuairisg} derived for this RR Lyrae, namely, 0.4034 days. 
Folding the observed light curve with our period and folding with the longer
period of 0.4034 days (see Fig.~\ref{foldrr}) both give reasonably
smooth curves. Thus, we cannot exclude that the period derived by
\citet{tuairisg} might be the correct one for this RR Lyrae.  

V14 and the previously known RR Lyrae V137 in Clement et al.\ (2001) 
match within $1\farcs8$, but the periods do not agree with each
other. We derived a period of 0.2997 days, which is smaller than the
0.3520 day period \citet{tuairisg} derived. The fit to the observed
light curve is better with our shorter period than with a 0.3520 day
period. Also, the folded light curves suggest that 0.3520 days is not
the correct period for this variable star.  

V27 shows a long-term variability with a period of 0.3531 days, which
agrees very well with the known RR Lyrae V160 in Clement et al.\
(2001), for which \citet{tuairisg} derived a period of 0.3529 days. The
coordinates of V27 and Tuairisg's V160 match within 
$2\arcsec$. NUV information is available for V27, and its location
in our FUV--NUV CMD is above the ZAMS and close to the ZAHB. This
agrees well with the expected location of RR Lyrae stars.

V28 is within $2\arcsec$ from the known RR Lyrae V135 (Clement et al.,
2001). We derived a period of 0.3615 days for this variable star,
which agrees very well with the 0.3619 day period obtained by
\citet{tuairisg}. 

\subsection{Newly detected RR Lyrae: V12, V16, V17, V19, V23, V29,
  V32, V34, V35} 
\label{rrlyraecand}

The coordinates of our RR Lyrae candidates V12, V16, V17, V19, V23, V29,
V32, V34 and V35 do not match those of any known RR Lyrae stars within
$2\arcsec$. All of these sources have FUV$_{STMAG} =$ 18.5 -- 21.3
mag and brightness variations of about 2 -- 4 mag on a timescale of 0.2
-- 0.6 days (see Table~\ref{periods1}). Except for V35, NUV
information is available for all of them, and V12 and V32 are located
between the BS and ZAHB sequence in our CMD in Fig.~\ref{cmdps},
whereas V16, V17, V23, V29, and V34 lie more or less exactly on the
ZAHB track. V19 is somewhat fainter and redder than most of the other
RR Lyrae candidates and is located above the faint end of the
ZAHB. The positions of all these variable sources agree with the 
expected location of RR Lyrae stars in the FUV--NUV CMD.  

All of the above-mentioned variables can be called RR Lyrae with good
confidence, as their periods, amplitudes, and location in
the CMD agree with what we expect for this type of variable. However,
V12 shows a larger scatter in its light curve than the other RR Lyrae,
and its location in the CMD is towards fainter FUV magnitudes. We
call this variable a suspected RR Lyrae.

The reason why these RR Lyrae were not found in previous studies could
be that optical surveys suffer from severe crowding that increases
towards the cluster core. Also, the FUV amplitudes for RR Lyrae are
more significant than the optical ones, making them easy to
detect. However, we point out that our variability census is not
complete as we only consider bright variables with large FUV
amplitudes. Variables that either are FUV fainter than 22 mag
or have smaller amplitudes, see Fig.~\ref{varsigma}, are not further
discussed in the present paper.        

\subsection{Previously Suspected Dwarf Novae: HCV2005-A (CV\,1) and HCV2005-B}
\label{dn}
 
DNe are CVs that show multiple outbursts ranging in
(optical) brightness from 2 to 5 mag. The outburst intervals
are quasi-periodic but can range from days to decades. The duration of
an outburst is typically days to weeks. These outbursts are thought to
be due to the release of gravitational energy, which results from an
instability in the accretion disk.  

In addition to the two LMXBs AC\,211 and M\,15 X-2,
\citet{hannikainen} found four faint X-ray sources within the inner 
$50 \arcsec$ of M\,15. Two of these sources are within our field of
view, namely, HCV\,2005-A and HCV\,2005-B (see Table~\ref{dntab}).

\begin{figure}
\centerline{
\includegraphics[width=5cm]{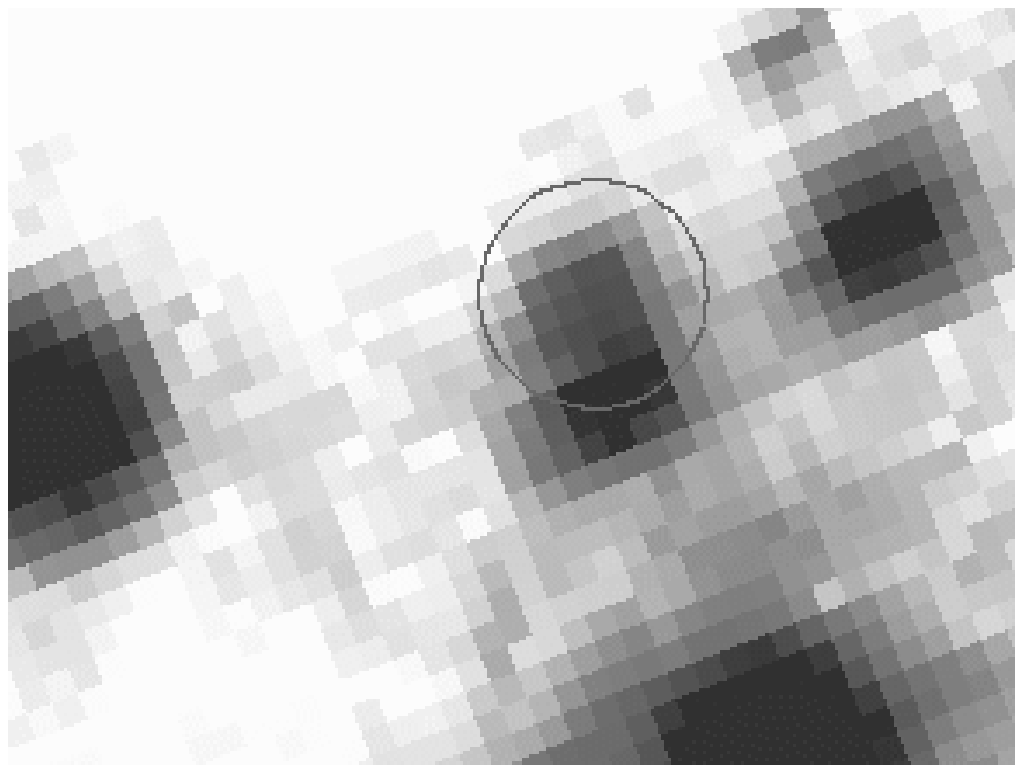}
\includegraphics[width=5cm]{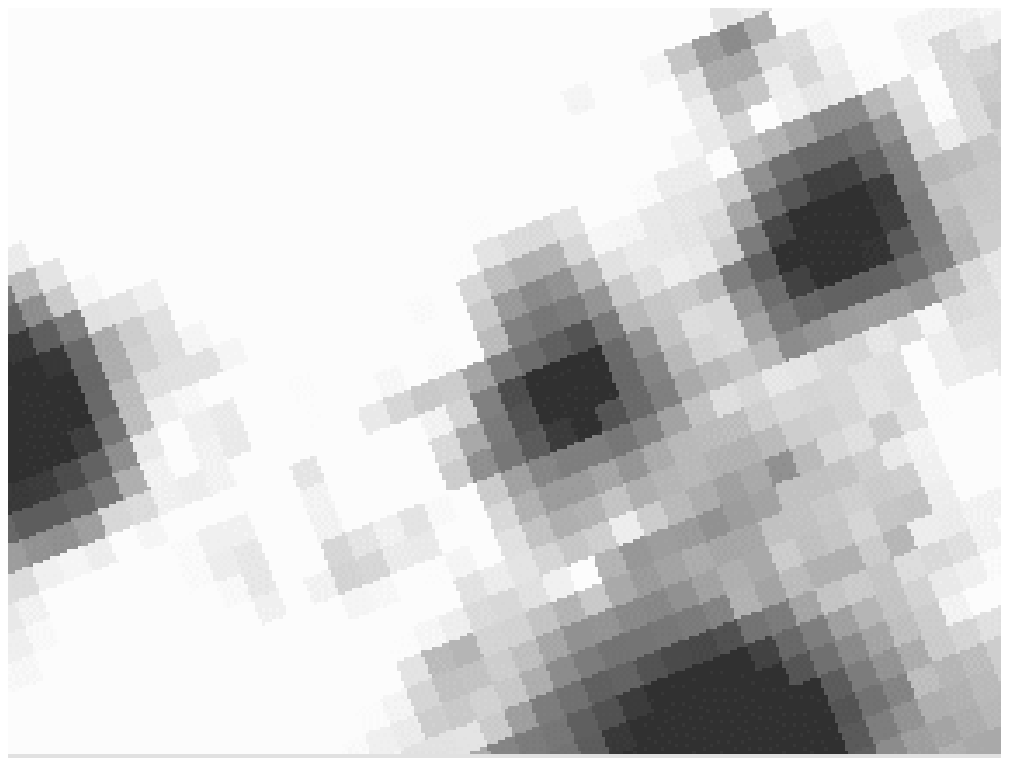}
\includegraphics[width=5cm]{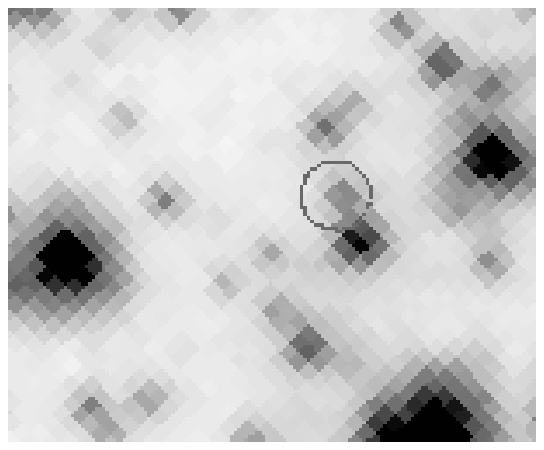}
}
\caption{\label{dn_cv1} Close-up of the DN CV\,1 or HCV\,2005-A that
 was in outburst only in our fourth FUV observing epoch (left) but
 was in quiescence during all other FUV observing periods
 (middle). The right panel shows a close-up in the NUV. Note 
 that we used the same orientation for these finding 
 charts as \citet{hannikainen} and \citet{charles}; i.e., north is to
 the upper left corner and east to the lower left. CV\,1 is marked with
 a circle and is located at $\alpha = 21^{h} 29^{m} 58.275^{s}$,
 $\delta = 12^{\circ} 10\arcmin 00\farcs1$. The field of view of these
 close-ups is $\approx 1\arcsec \times 0\farcs7$.}    
\end{figure}

\citet{hannikainen} suggested that HCV\,2005-A is a probable DN. This
source is the same as the DN CV\,1 in \citet{shara}. This object does not
appear in our initial list of FUV sources and was not noticed when
we inspected the FUV master image by eye. The reason for this is that
it is very faint and so close to the nearest bright star that the
two blend into each other and could not be discriminated. As
\citet{shara} pointed out, the DN is in quiescence during their FUV
observations and too faint to be detected. \citet{shara} used the same
data set as we do but could only use a part of the FUV epochs as the
remaining observations were not yet available at that time. In fact,
CV\,1 is in outburst in our fourth observing epoch, and so we were able
to identify this DN (see Fig.~\ref{dn_cv1}). This source appears to be
undetectable in the WFPC2 1994 April data, on which the \citet{vandermarel}
catalog is based, whereas the WFPC2 data taken in 1994 October show
a bright source (see also Hannikainen et al.\ 2005). 

\begin{figure}
\centerline{
\includegraphics[height=5cm]{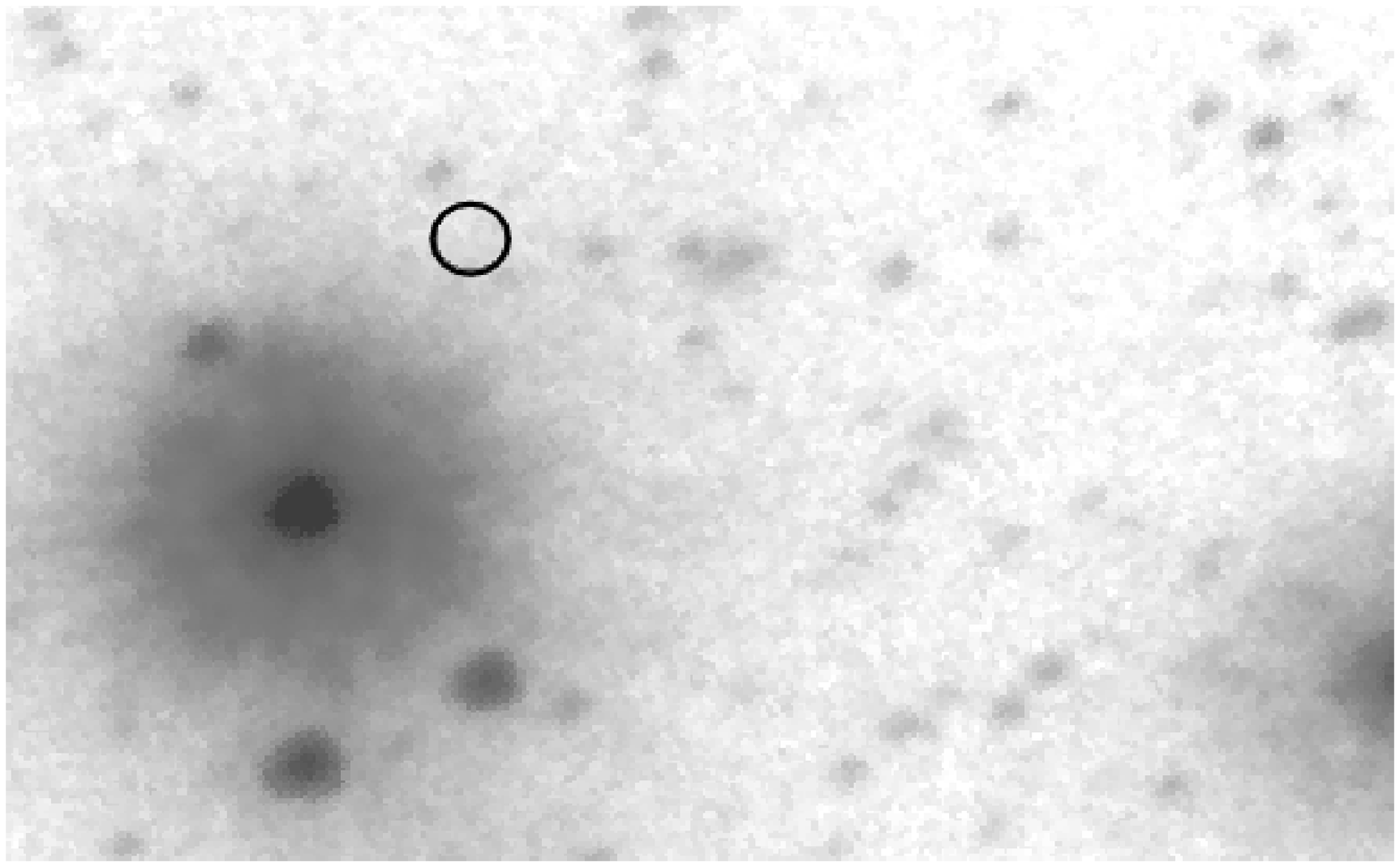}
\includegraphics[height=5cm]{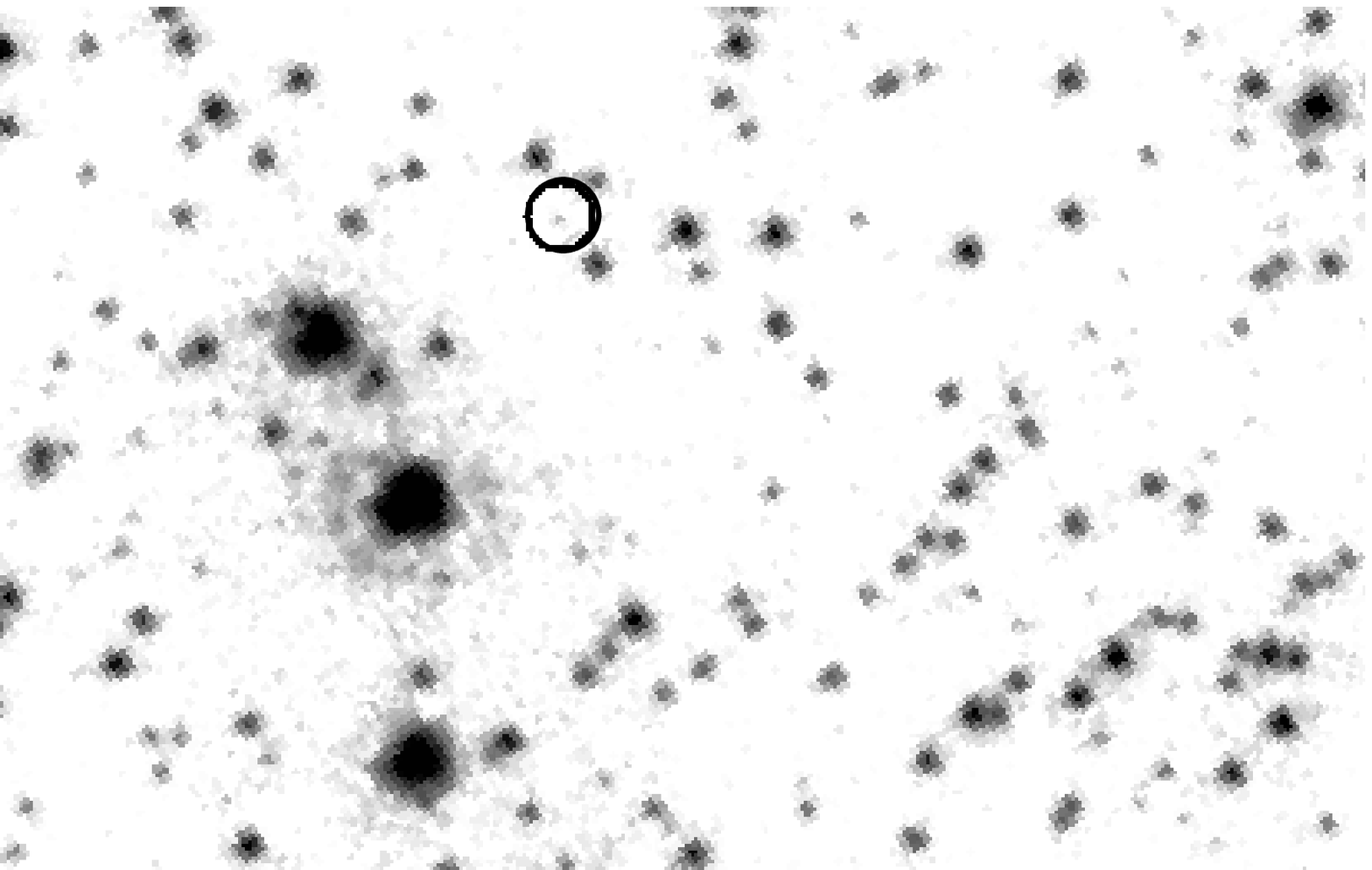}
}
\caption{\label{dn_hcv} Same as Fig.~\ref{dn_cv1}, but for the DN
 HCV\,2005-B. North is to the top left and east to the bottom
 left. HCV\,2005-B appears as a faint NUV source (right), but is too
 faint to be detected in our FUV data (left).The field of view of
 these close-ups is $\approx 4\farcs5 \times 3\arcsec$. The location
 of HCV\,2005-B in our image is at $\alpha = 21^{h} 29^{m} 58.226^{s}$,
 $\delta = 12^{\circ} 10\arcmin 11\farcs4$.}    
\end{figure}

The second X-ray source HCV\,2005-B is a faint NUV = 21.36 mag object
that appears only in our NUV images and is too faint to be detected
in our FUV data in any epoch (Fig.~\ref{dn_hcv}). 
\citet{hannikainen} suggested that this source is most likely a DN
but might also be a quiescent soft X-ray binary. Unfortunately, we
cannot shed more light on the true nature of this object.  
 
\begin{table} 
\begin{center}
\caption{\label{dntab} Cartesian coordinates in the NUV master image
  and $NUV$ magnitudes for the two DNe HCV\,2005-A (or CV\,1) and HCV\,2005-B.} 
\begin{tabular}{ccccc}
name      & x         & y        & NUV  & $\Delta$ NUV \\
\hline
HCV\,2005-A & 744.462   & 722.678  & 18.913 & 0.031        \\
HCV\,2005-B & 771.660   & 1180.19  & 21.361 & 0.051        \\
\hline
\end{tabular}
\end{center}
\end{table}

\subsection{Other Cataclysmic Variables: V7, V11, V15, V39, V40, V41}
\label{cvcand} 

The light curve of our variable V7 shows brightness
variations of a few tenths of a magnitude on short timescales of
minutes. CVs can show sizeable variability on a wide range of
timescales, and the variation we observe here might be the so-called
flickering, which is a random, aperiodic variability seen on
timescales of seconds to several minutes and typical for CVs
(e.g. Bruch 1992).   
The location of this variable source in the FUV--NUV CMD is close to
the WD cooling sequence, which makes this source an excellent CV
candidate. V11, V39, and V41 also are very good CV candidates as their
light curves show flickering and they are all located in the gap region
in our CMD, see Fig.~\ref{cmdps}. 

V15 and V40 show similar small time-scale variations. As no NUV
information is available for those two sources, we call these
variables suspected CVs.

\subsection{Cepheids: V10}
\label{cepheids}

Cepheids are radially pulsating, bright variables with periods of
$\approx 1 - 135$ days and variability amplitudes from several 
hundredths to 2 mag in $V$. The amplitude in $B$ is larger
and increases towards the FUV. 
Traditionally, both Delta Cep and W Vir stars are called Cepheids as
it is often impossible to discriminate between them on the basis of
the light curves for periods in the range 3 -- 10 days. However, these
are distinct groups of entirely different objects in different
evolutionary stages. Delta Cep stars, the classical Cepheids, 
are relatively young objects belonging to the young disk
population. They can be found in open clusters and have just evolved
into the instability strip of the Hertzsprung-Russell diagram. W
Virginis variables, on the other hand, belong to the old disk
population and can be found in GCs and at high Galactic
latitudes. They show periods of approximately 0.8 -- 35 days and
amplitudes from 0.3 to 1.2 mag in $V$. The period-luminosity
relations are different for Delta Cep and W Virginis variables. For
more details, see \citet{GCVS}.

The coordinates of V10 closely match (within $0\farcs2$) with the
Cepheid FP V16, and its period of 1.62 days also agrees with this
variable source being a Cepheid star. \citet{tuairisg} found a shorter
period of only 1.2411 days for this object, but we note that their
estimate is close to one of our aliases in our periodogram. The
location of V10 in our CMD is above the ZAHB sequence.

\subsection{Other possible Cepheid candidates: V1, V13, V18, V30, V31, V36}
\label{cephcand}

The most striking feature in the light curve of our variable 
V1 is that it brightens by $\approx 3$ mag in the fifth observing
epoch. This might indicate an outburst, as in DNe. On the other hand,
V1 might be a variable with a long-term period, possibly 2.6 days,
which might suggest that this source is a Cepheid. We
do not have additional NUV information, as this star is located
outside the NUV field of view. However, the counterpart to V1 in
\citet{vandermarel} is a bright source located on the BHB in the
optical CMD (Fig.~\ref{cmd_opt}). This suggests that this
variable is a Cepheid.  

V13 is $\approx 1 - 2$ mag brighter in the fifth and sixth
observing epochs compared to the remaining epochs. This again might indicate
an outburst as in DNe, or it might indicate a long-term variability of
possibly 3.4 days, which is consistent with a Cepheid periodicity. Its
location in our FUV--NUV CMD is above the ZAMS and close to the
ZAHB. As CV candidates are expected in the region between the ZAMS and
the WD cooling sequence, we suggest that V13 is a Cepheid candidate
rather than a CV candidate.  
The same is true for our variable sources V18, V30, V31, and
V36. All sources show an increase in brightness of about 2 mag in at
least one observing epoch, which might be indicative of a DN
outburst. However, periods of 1.4 days for V18, 4.5 days for V36, and
3.8 days for V30 and V31 fit to the corresponding light curves which
might suggest that these stars are Cepheids. V18 and V36 are located above
the ZAHB in our FUV--NUV CMD, see Fig.~\ref{cmdps}, whereas V30 and
V31 are located above the ZAMS and below the ZAHB, but close to the
location of the previously known Cepheid (our V10). The location of
all these variables agrees with the expected location of Cepheids in
our CMD.

We caution that our variable classification is a tentative one, based
on the derived periods and the photometric quantities of the variables.
This is especially difficult for our Cepheid candidates, as
the long periods are not evenly or completely covered by our
data set. This is reflected in Fig.~\ref{fold}, which shows the folded
light curves for all our variables. As can be seen, there are
quite large gaps in the folded light curves for our Cepheid candidates,
rendering their classification somewhat uncertain. However, except for
V31, all our Cepheid candidates have optical counterparts, and these
are all located along the BHB in the optical CMD, the expected
location for Cepheid variables. 

\subsection{Low-mass X-ray binaries: V21/AC\,211 and M\,15 X-2}
\label{lmxb}

V21 is the known LMXB 4U\,2129+12 or AC\,211. It is not only the
brightest variable but also the brightest FUV source in our
catalog. This source has a period of 17.1 hr (Ilovaisky et
al.~1993) and is one of the optically brightest LMXBs known. The
light curve of V21/AC\,211 is presented in Fig.~\ref{lcurve2} and seems
to indicate an eclipse during the sixth observing
epoch. Unfortunately, our own data coverage is not good enough to
derive a reliable period for this source.  

As mentioned earlier, the second LMXB in this cluster, M\,15 X-2, does
not show a large $\sigma_{mean}$ in Fig.~\ref{varsigma}. We refer the
reader to \citet{dieball2} for a thorough discussion of this UCXB. For
the sake of completeness, we include M\,15 X-2 in Table~\ref{periods2}
and mark this source in our CMD in Fig.~\ref{cmdps}.  

\subsection{SX Phoenicis: V33}
\label{sxphoen}

SX Phoenicis stars are pulsating BSs with short periods of less
than 0.1 days \citep{jeon}. They are found in the old disk population
and in GCs, and show variability amplitudes of up to 0.7
mag in $V$.  

V33 shows large brightness variation of $\approx 3$ mag with a
period of 1.32 hr. A sine wave with a corresponding period fits
the observed light curve well, see Fig.~\ref{lcurve3}. The
coordinates for V33 agree within $0\farcs8$ with the known SX
Phoenicis star VGG 10041, which is ZK\,62 in \citet{kravtsov}, who derived a
period of 1.248 hr for this variable star. The FUV amplitude
is much larger than the optical one. The same effect is known for RR
Lyrae stars and Cepheids. Large FUV amplitudes are to be expected
since the amplitudes of the pulsations increase towards the
FUV \citep{downes, wheatley}.  

\subsection{Another SX Phoenicis candidate: V22}
\label{sxphoencand}

Our variable V22 shows a 2 mag brightness variation with a
period of 1.06 hr. Its location in our CMD, see Fig.~\ref{cmdps}, is
slightly above the ZAMS and BS track. Its characteristics suggest that
it might be an SX Phoenicis star.

\subsection{Other suspected variables: V5, V6, V20, V24, V25, V26, V37, V38}
\label{interact}

The light curve of V5 shows a small-amplitude scatter, but the most
striking feature is the rise in brightness at the end of the first
observing epoch, which might correspond to an egress from an
eclipse, or the beginning of an outburst as in DNe. The source is faint
in the second, third, fourth, and sixth observing epochs, but brighter
in the fifth observing epoch, which fits the characteristics of a
DN. The source is located between the BS and ZAHB sequence in our
FUV--NUV CMD, see Fig.~\ref{cmdps}. However, its optical counterpart
is located on the BHB in the optical CMD. If this source is indeed a
DN in M15, its optical counterpart is unusually bright. Further
observations are required to shed more light on this variable source.

The light curve of V6 shows a considerable drop in brightness in the
second, third, and fifth observing epochs. The characteristics of the
light curve match that of a DN. In this case, V6 seems to be in
outburst during the first, fourth, and sixth observing
epochs. This would imply an unusually high duty cycle of 70\% of the
time coverage. Unfortunately, this star is located outside the NUV
field of view.   
However, we do have an optical counterpart to this source, which
is a rather unexpected one: it is one of the brightest optical
sources, and its location in the optical CMD is blueward of the tip
of the RG branch at $V=12.953$ and $B-V = 0.737$ mag (see
Fig.~\ref{cmd_opt}). Its optical magnitude is far too bright to agree
with a DN at the distance of M\,15. A possible explanation is that the
optical counterpart is a ``mismatch'' (although the FUV and optical
coordinates agree within $0\farcs0165$),
and the ``true'' optical source might be an undetected, faint source
in nearly the same line of sight as the bright one. Another, and
perhaps more likely, explanation is that this source is a variable
field star that happens to be in our line of sight. If it is a field
DN, it would be much closer to us than M\,15, i.e.\ at least 400 pc
assuming an absolute $M_V = 5$ mag for a DN in outburst. 

Our variable V20 shows flickering and in addition reveals some
long-term variability. It is brightest during our first, third, fourth,
and fifth observing epochs. This might indicate outburst activity,
as in DNe. It is also possible that V20 is a CV of VY Scl type, which
show sudden drops in brightness. If this is true, V20 might be in a
low state in the second and sixth observing epochs. On the other hand,
its light curve could also be explained with a Cepheid-type long-term
periodicity of 1.5 days. This suspected variable is a bright source
and is located just below the ZAHB in our CMD. Based on its light curve
and location in the CMD alone, we cannot decide whether this source
might be a bright CV or a bright Cepheid, and further observations are
required to decide on the nature of this variable.

V24 is located above the clump of MS stars and RGs and on the
lower (fainter) part of the ZAHB in our FUV--NUV CMD. From its location
in the FUV--NUV CMD, it might well be an FUV fainter but redder BS
or a BHB star. However, the optical counterpart to this source is
located on the BHB in the optical CMD. This suggests that this source
might be a binary or pulsating HB star.  

The light curve of V25 shows some small-scale brightness variation on
short time-scales, possibly flickering. However, in addition to the
flickering, a long-term variability seems to be superposed, which can
be seen in the first observing epoch where the source is $\approx 2$
mag fainter towards its end. The location of V25 in our CMD is on
the lower, fainter part of the ZAHB. This source is difficult to
classify, and it might be a binary or pulsating BS or HB star.

V26, V37, and V38 show flickering with a few tenths of a magnitude
variation on short time-scales. Their location in the FUV--NUV CMD is
close to the ZAMS and they might be either bright CVs or binary or 
pulsating BSs. Only V37 has an optical counterpart, but its optical
magnitudes agree again with a BS or a bright CV. 

\subsection{Variability Census and its Implications}

We found 41 sources that show strong signs of variability. Based
on the appearance of their light curve (Figs.~\ref{lcurve1} to
\ref{lcurve3}), their period, and their location in the FUV--NUV and
in the optical CMDs (Figs.~\ref{cmd} and \ref{cmd_opt}), we suggest
classifications for our variable sources. 
We compared the coordinates and periods, if available, of our variable
sources with known variables in the core region of M\,15. In total,
we found 17 RR Lyrae stars (4 of which are known), 6 CV
candidates, 1 known SX Phoenicis star, 1 SX Phoenicis
candidate, 1 previously known Cepheid and 6 further candidates,
and the known LMXB AC\,211. The 8 remaining suspected variables
cannot be classified clearly, but most of them might be CVs or
pulsating/binary BSs.  

Note that only six out of our gap objects show strong signs of
variability. However, the remaining gap objects might also be
variable, but they are either fainter or have smaller amplitudes and
thus were not detected in our variability study, see
Fig.~\ref{varsigma}. (We note again that, for 
example, the UCXB M\,15 X-2 was not detected as a strongly variable
source in this study, as it shows a modulation with only 0.062 mag
semi-amplitude; see Dieball et al.\ 2005b.) We marked the gap objects
that we selected based 
on our FUV--NUV CMD as red filled circles in Fig.~\ref{varsigma}. The
green filled circles denote the two sources that do not have an NUV
counterpart, and thus do not appear in our FUV--NUV CMD, but are
likely CVs because of 
their variability. There are $\approx 60$ gap objects/CV candidates. 
Typical short-timescale variability for a CV is flickering with
amplitudes of a few tenths of a magnitude. As can be seen from
Fig.~\ref{varsigma}, there are 25 gap sources that show a
$\sigma_{mean} > 0.5$ mag, but most of these are located in the bulge
of faint sources at FUV$_{mean} > 22$ mag and thus were not considered
in the above study. We consider these sources good CV candidates. This
number agrees with our previous estimate, see Sect.~\ref{cv}.

\section{Summary}
\label{summary}

We have analyzed deep FUV {\it HST} ACS observations of the core of
the metal-poor GC M\,15. Based on our FUV and NUV
data, we constructed a CMD in which various stellar populations are
present. Our CMD is deep enough that MS stars and RGs show
up. The MS turnoff is visible at FUV $\simeq 23.5$ mag and FUV--NUV 
$\simeq 3$ mag, and the MS stars and RGs form a prominent data
clump that extends at least 2 mag below the MS turnoff. 
As such, this is the deepest FUV CMD presented so far. A line of
$\approx 70$ BS candidates and a well-defined sequence of $\approx
130$ HB stars are present in our CMD. BHB stars are located along the
well-defined HB sequence, whereas EHB stars are clustered within a
bright data clump. Approximately 30 WD candidates are found close to
the WD cooling sequence, which is well within the expected number of
WDs in our field of view down to FUV $\simeq 22.5$ mag. The region
between the WD cooling sequence and the ZAMS and BS track is occupied
by $\approx 60$ objects. Theoretical considerations suggest that at
least some of these are CVs. We caution that these numbers should not
be taken as exact, since the discrimination between the various zones
in our CMD is difficult. The cumulative radial distributions of the stellar
populations in our FUV--NUV CMD suggest that gap sources and BS
candidates are the most centrally concentrated populations. This
might be the expected effect of mass segregation; i.e., more massive
stars and binaries like CVs and BSs sink towards the cluster
core. Alternatively, it might reflect the preferred birthplace of such
objects, as CVs and BSs are thought to be dynamically formed
and as such are preferentially found in the dense cluster core. We
found that also EHB stars are more concentrated in the core region than BHB
stars. This might indicate that at least some EHB stars also have a
dynamical origin. 

We searched for variability among our FUV catalog stars and found
41 variable sources. They are distributed over the entire CMD,
but most of them are located around either the ZAHB or the
ZAMS. Periods could be derived for 27 variables. Based on their
variability properties and their location in the FUV--NUV 
CMD, we suggest classifications for 33 variables. Seven
variable sources could be identified with previously known variables.
 
In total, we found four known RR Lyrae stars, 13 additional RR Lyrae
candidates, six CV candidates, two SX Phoenicis stars (of which one
is known), seven Cepheids (one previously known), and the known LMXB
AC\,211. The eight remaining suspected variables 
cannot be classified clearly, but most of them might be CVs or
pulsating/binary BSs. We found that the brightness variations for RR
Lyrae stars, Cepheids, and SX Phoenicis stars are a few magnitudes
in the FUV and thus much larger than in the optical. This is to be
expected since the amplitudes of the pulsations increase towards the
FUV \citep{downes, wheatley}. However, this is the first time such
large-scale FUV amplitudes have been observed for SX Phoenicis stars. 
Note that out of our 60 gap objects/CV candidates, only 6 show strong
signs of variability. However, 25 of the gap sources show
$\sigma_{mean} \ge 0.5$ mag. We consider at least these to be good CV
candidates. This number agrees with theoretical predictions.

All previously known X-ray sources and DNe that are located in our
field of view are detected in our UV study. The LMXB AC\,211 is the
brightest FUV source in our sample, and also the UCXB M\,15 X-2 is
among the brightest sources. HCV\,2005-A and HCV\,2005-B
\citep{hannikainen} are both detected in our NUV data set, although
HCV\,2005-B appears only as a faint NUV source and is not detected in 
the FUV. HCV\,2005-A or CV\,1 \citep{shara} is in outburst in our
fourth FUV observing epoch, but otherwise it is too faint to be detected
in the FUV. 

Overall, the results of our study confirm that FUV observations are
particularly well suited in studying hot, and especially dynamically-formed,
stellar populations like CVs, BSs, and X-ray binaries in the cores of
GCs. They are also a powerful tool in detecting new variable stars in
the dense core regions of GCs where optical studies are hampered by
the immense crowding. In addition, the large FUV amplitudes of
especially RR Lyrae stars, Cepheids, and SX Phoenicis stars make them easily
detectable as variables in this wave band. 

All clusters that have been studied so far in the FUV, namely, 47\,Tuc,
NGC\,2808, and M\,15, revealed large numbers of gap objects that were
more or less consistent with theoretical predictions for dynamically
formed CVs. It is interesting to directly compare NGC\,2808 and M\,15,
as both clusters now have been studied in both the FUV and NUV,
with a full coverage of the cluster cores. Table~\ref{gc} lists the
cluster parameters taken from \citet{harris} and the number of gap
objects and BS and WD candidates found in the clusters. As can be
seen, slightly more gap sources were detected in NGC\,2808 than in
M\,15, whereas more BS candidates are present in M\,15 than in
NGC\,2808. Evidence for a concentration of the more massive CVs and
BSs is more evident in NGC\,2808 (see Dieball et al.~2005a, their Fig.~4),
which is somewhat contrary to the longer relaxation time for the
cluster's core (see Table~\ref{gc}). However, we caution that these
are still small number statistics, so their difference is not
particularly significant. We are currently working on data of more
clusters. Our current sample with just two GCs is too small to draw
any conclusions about the impact of a difference in the binary
populations on the cluster evolution. However, the large number of
exotica and massive binaries found in the cores of these clusters
agrees with theoretical expectations for evolved clusters.  

\begin{table}[hb!]
\begin{center}
\caption{\label{gc} Number of gap sources and WD and BS candidates detected in
  GC. The cluster metallicity and the logarithmic core relaxation time
  log(tc) and halfmass relaxation time log(th) are taken from \citet{harris}.} 
\begin{tabular}{lcccccc}
\tableline
name      & gap & WDs & BS & [Fe/H] & lg(tc) & lg(th) \\
\tableline
NGC\,2808 & 60  & 40  & 61 & -1.15  & 8.30   & 9.13 \\
M\,15     & 57  & 28  & 69 & -2.26  & 7.02   & 9.35 \\
\tableline
\end{tabular}
\end{center}
\end{table}

\acknowledgments

We are grateful to an anonymous referee for a valuable discussion that
helped us to improve this paper. DH acknowledges the Academy of Finland.
This work was supported by NASA through grant GO-9792 from the Space
Telescope Science Institute, which is operated by AURA, Inc., under NASA
contract NAS5-26555.

\end{document}